\documentclass{aastex631}
\usepackage{xcolor}

\definecolor{mathpurple}{RGB}{164,98,155}
\definecolor{f1}{RGB}{95,130,179}
\definecolor{f2}{RGB}{210,146,49}
\definecolor{f3}{RGB}{144,175,60}
\definecolor{darkorange}{RGB}{195,110,39}

\shorttitle{Seismic Wave Transmission on Granular Asteroids}
\shortauthors{S{\'a}nchez et al.}
\graphicspath{{./}{/}}

\begin{document}

\title{Transmission of a Seismic Wave generated by impacts on Granular Asteroids}

\correspondingauthor{Paul S{\'a}nchez}
\email{diego.sanchez-lana@colorado.edu}

\author[0000-0003-3610-5480]{Paul S{\'a}nchez}
\affiliation{Colorado Center for Astrodynamics Research \\
University of Colorado Boulder\\
3775 Discovery Dr \\
Boulder, CO 80303, USA}

\author{Daniel J.~Scheeres}
\affiliation{Ann \& H.J. Smead Department of Aerospace Engineering Sciences\\
University of Colorado Boulder\\
3775 Discovery Dr \\
Boulder, CO 80303, USA}

\author{Alice C. Quillen}
\affiliation{Department of Physics and Astronomy \\
University of Rochester\\
Rochester, NY 14627, USA}


%
%
%



\begin{abstract}

In this paper we use a Soft-Sphere Discrete Element method code to simulate the transmission and study the attenuation of a seismic wave.  Then, we apply our findings to the different space missions that have had to touch the surface of different small bodies.  Additionally, we do the same in regards to the seismic wave generated by the hypervelocity impacts produced by the DART and Hayabusa2 missions once the shock wave  transforms into a seismic wave.

We find that even at very low pressures, such as those present in the interior of asteroids, the seismic wave speed can still be on the order of hundreds of m/s depending on the velocity of the impact that produces the wave.  As expected from experimental measurements, our results show that wave velocity is directly dependent on $P^{1/6}$, where $P$ is the total pressure (confining pressure plus wave induced pressure).  Regardless of the pressure of the system and the velocity of the impact (in the investigated range), energy dissipation is extremely high.  These results provide us with a way to anticipate the extent to which a seismic wave could have been capable of moving some small particles on the surface of a small body upon contact with a spacecraft.

Additionally, this rapid energy dissipation would imply that even hypervelocity impacts should perturb only the external layer of a self-gravitating aggregate on which segregation and other phenomena could take place.  This would in turn produce a layered structure of which some evidence has been observed.

\end{abstract}



\section{Introduction} \label{sec:intro}

The different space missions that have visited some of the many asteroids in the Solar System in the the last decades have made it clear that these small bodies are gravitational aggregates held together by gravitational, cohesive and adhesive forces \citep{science_fujiwara, science_yano, thomas_eros_craters, richardson2009, scheeres2010, sanchez2014, sanchez2016}.  In-situ observations of asteroids Eros, Itokawa, Ryugu and Bennu as well as  thermal and radar observations of other asteroids have shown that they can be covered with grains with a size distribution that spans from microns to tens of meters \citep{itokawa_boulders,itokawa_miyamoto}. Images taken by the OSIRIS-REx and Hayabusa2 missions, while their spacecrafts were in orbit, also show or imply the presence of dust (10-500$\mu$m) on the surface of asteroids Bennu and Ryugu respectively \citep{dellagiustina2019,jiang2020,tachibana2022}.  Additionally, the current understanding points to asteroids that have evolved collisionally to result in the bodies that we can observe today; however, most of the impacts involved are non-disruptive.  If this is so, small, non-disruptive impacts should be common and energetic enough to produce seismic  waves and possibly even particle size segregation on asteroid surfaces \citep{sanchez-lpsc2010, perera2016, wright2020}.

The study of sound waves in granular matter is not new, with studies going back to more than 80 years ago \citep{hara1935}.  However, experimental studies deal with pressures that start at tens of kilopascals; orders of magnitude higher than the typical internal pressure of a small asteroid.  In spite of this, if one thing is clear from a theoretical point of view is that sound speed and pressure are intimately related \citep{walton1977,digby1981}.  If we take the tendency of the experimental results \citep{goddard1990}, this would mean that impacts at a few hundreds of meters per second on a small planetary body should be treated as supersonic.

The study of sound transmission in granular matter and porous materials, has been carried out through experimental, theoretical and numerical means, but generally under terrestrial conditions.  The theory that has emerged is called Effective Medium Theory (EMT) and is based in considering the media as elastic, removing so the difficulties presented by particle size and shape.  Based on this assumption, it is possible to calculate, from first principles, the bulk ($K$) and sheer ($\mu$) moduli and from them, the speed of sound.  This theory, based on the Hertzian laws for contacts, predicts a sound velocity that has a $P^{\frac{1}{6}}$ dependence on pressure ($P$), whereas the experiments have found this is true only for high enough pressures and that at low pressures the dependency changes to $P^{\frac{1}{4}}$ \citep{makse1999,velicky2002} .

Seismic waves are waves of energy that can be caused by many different means.  There are several different kinds of seismic waves, and they all move in different ways. The two main types of waves are body waves and surface waves.  As defined, body waves travel through the interior of a planetary body and surface waves, only through its surface.  In turn, body waves can be divided into two kinds: primary (pressure) waves (P waves) and secondary (shear) waves (S waves).  

More recently, other researchers have also studied the behaviour of seismic waves under low confining pressure.  Simulations in 2D idealised systems in which a granular bed of frictionless grains is continually compressed by a piston at a constant speed have been carried out to understand the behaviour of seismic waves in these conditions \citep{gomez2012}.  It was found that the speed of sound is dependent on the speed of the piston and the authors showed that a theory based on the micro-compression of the grains could explain their findings.  A subsequent study \citep{wildenberg2013} showed experimental results had a good correlation with the theory developed previously by \cite{gomez2012}.  Unfortunately, experiments in 3D systems are still rare \citep{tell2020,zeng2007}, but some experimental apparatuses  have been developed with this very objective in mind \citep{aumaitre2018}.  It is also important to note, that most experiments have been carried out in the kPa-MPa regime, whereas the pressure at the interior of small asteroids are orders of magnitude smaller.  In this research will study seismic wave transmission at pressure levels close to those found in asteroids and up to tens of kPa to overlap with the existing measurements. 

This being said, the study of seismic wave transmission in granular systems under low confining pressure has several potential applications in the context of asteroid research and space exploration.  This would include modelling seismic shaking of granular asteroids, possible future seismology experiments on asteroids, landing, roving, anchoring and sampling on their surfaces, and improving our understanding of the mechanics of microgravity aggregates.

\section{Research Methodology} 

For this research we will use a Soft-Sphere Discrete Element Method (SSDEM) code.  This method has been used by other researchers in the field, specifically for research on sound wave transmission in granular media, with great success.  At the moment, the code we have developed is able to simulate a self-gravitating asteroid size aggregate \citep{sanchez2014, sanchez2016}; however, this approach presents two problems, one is numerical and the second is physical.  Numerically, the number of particles that can be simulated, to make the simulation time accessible, is limited and also their size distribution.  This means that the simulations will be low-resolution.  Additionally, due to the nature of a self-gravitating body, there will be a pressure gradient that will make it difficult to establish a well defined relationship between confining pressure and wave speed.

A second option, given that we have a parameter space with at least 3 variables (impact speed, particle-particle tensile strength and confining pressure) is to explore it in such a way that the parameters that characterise a small asteroid are within the examined region of our parameter space.  This would offer us a greater resolution and hopefully a better understanding of the phenomenon at hand and also how to relate our findings to the experimental findings for Earth bound systems.  In view of this, this is the approach we have decided to follow.  In a way, it will be similar to placing several small test boxes ($\gg$typical particle size) inside a non-monolithic, small planetary body and seeing how the wave is transmitted inside them.

The parameter space that is studied is chosen so that the parameter values that characterise a small granular asteroid are included in the studied region.   In the particular case at hand, parameters such as confining pressure ($P$), impact velocity ($v$) and particle-particle tensile strength ($\sigma_{yy}$) seem to be relevant.  The last one however will be left for a future report and set to zero given the extremely low cohesion found in asteroids Bennu and Ryugu \citep{roberts2021}.  Theory tells us that $P$ is directly related to wave speed; it has been proposed that small impacts on granular asteroid could be responsible for particle size segregation; tensile strength has been proposed as a major influence for the structural stability of small asteroids \citep{scheeres2010,sanchez2014}.  To explore these two parameters we  have set up a system comprising of a simple rectangular container, with horizontal periodic boundaries, filled with a number of grains on which pressure can be applied by a moving top (piston). Such a system would provide us with a much greater resolution and control than the simulation of an asteroid-size aggregate discussed above and so we will adopt it.

\section{Simulation Code and Setup}

The simulation program that is used for this research applies a Soft-Sphere Discrete Element Method (SSDEM) \citep{cundall1971, cundall1992}, implemented as a computational code (in house developed) to simulate a granular system \citep{bis, sanchez-lpsc2009, sanchez2011,sanchez2012}. For this particular case, the particles are modelled as perfect spheres with radii that follow a well determined size distribution.  In this method, two particles are said to be in contact when there is an overlap which, for spheres, takes place with the distance between any two particles is small than the sum of their radii.  Other shapes can also be modelled; however, the contact detection algorithms have to be much more sophisticated.  When a contact takes place, normal and tangential contact forces are calculated \citep{herr1}. The former is modelled by a Hertzian spring-dashpot system and is always repulsive, keeping the particles apart; the latter is modelled with a linear spring that satisfies the local Coulomb yield criterion.  The normal elastic force is modelled as
\begin{equation}
{\vec{\bf f}}_e= k_n\xi^{3/2}{\bf\hat n},
\label{hertz}
\end{equation}
the damping force as:
\begin{equation}
{\vec{\bf f}}_d=-\gamma_n\dot\xi{\bf\hat n},
\end{equation}
where $r_1$ and $r_2$ are the radii of the two particles in contact. Then, the total normal force on a particle is calculated as ${\vec{\bf f}}_n={\vec{\bf f}}_e+{\vec{\bf f}}_d$.  In these equations, $k_n$ is the elastic constant, $\xi$ is the overlap of the particles, $\gamma_n$ is the damping constant (related to the dashpot to simulate energy dissipation in a collision), $\dot\xi$ is the rate of deformation and ${\bf\hat n}$ is the vector joining the centres of the colliding particles. This dashpot models the energy dissipation that occurs during a real collision.  Note that linear, instead of Hertzian springs, can also be used in this simulation method; however, we have chosen to use the latter option due to our application.

The tangential component of the contact force models static and dynamic surface-surface friction. This is calculated by placing a linear spring attached to both particles at the contact point at the beginning of the collision \citep{herr1,silbert2001} and by producing a restoring frictional force ${\vec{\bf f}}_{t}$. The magnitude of the elongation of this tangential spring is truncated in order to satisfy the local Coulomb yield criterion $|{\vec{\bf f}}_t|\leq\mu_k |{\vec{\bf f}}_n|$, where $\mu_k$ is the coefficient of dynamic friction.
    
\begin{figure*}[h]
\begin{center}
\includegraphics{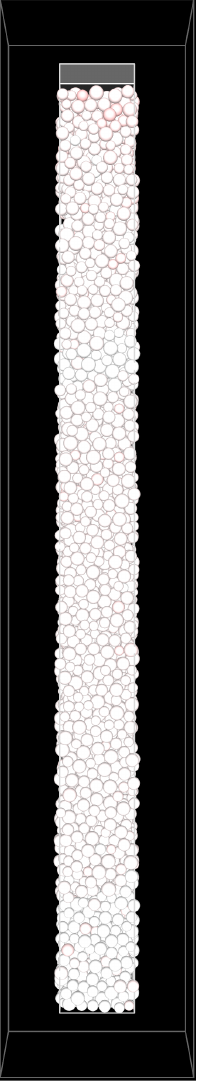}
\hspace{-2.5mm}
\includegraphics{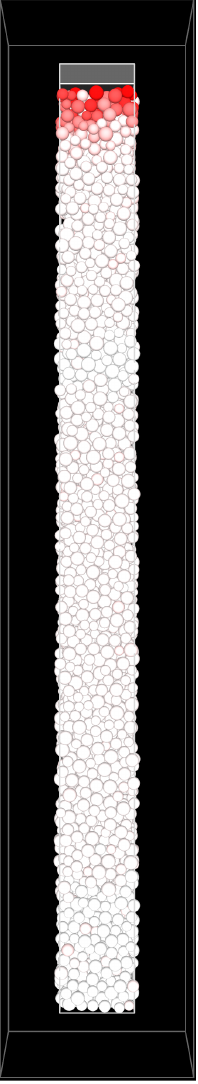}
\hspace{-2.5mm}
\includegraphics{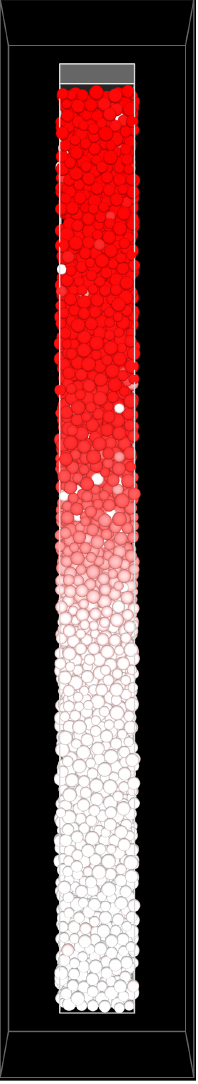}
\hspace{-2.5mm}
\includegraphics{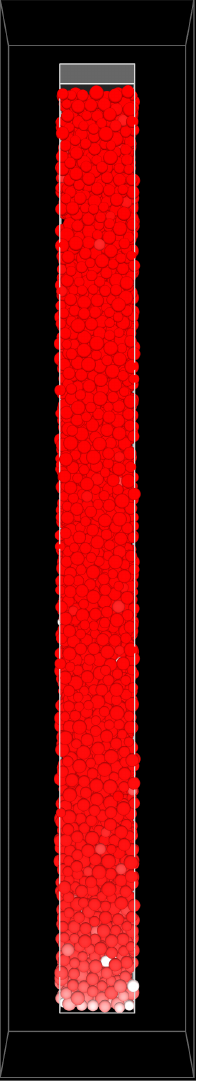}\\
\vspace{-1mm}
\hspace{-1.35mm}
\includegraphics[scale=0.7082]{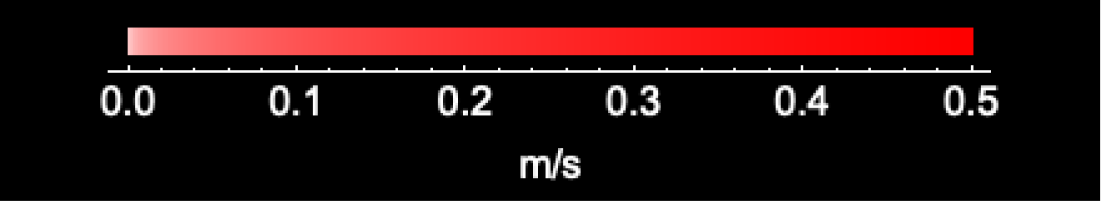}
\caption{Simulation setup and seismic wave transmission through a granular medium.  The redder the colour of a particle, the greater its speed in the vertical direction.  From the left to the right, t=0.5, 0.50005, 0.5021 and 0.5050 s.  The system is subjected to a confining pressure of 1 kPa and the piston is moving at a constant speed of 0.5 m/s.}
\label{sim-ini}
\end{center}
\end{figure*}

For this particular set of simulations, we have chosen to use material parameters close to those of basalt so that the results are relevant for asteroids.  We have chosen to simulate perfectly spherical grains in order to remove a parameter that would require a dedicated research effort on its own.  Particles will have a random close packing (RCP) to simulate a natural arrangement.  We will be using 3120 spherical particles, 2-3 cm in diameter and follow a uniform random distribution.  This size variation will avoid the formation of a crystalline structure.

Particle density is 3200 kg m$^{-3}$, the average density of LL chondrites (the low-iron, low metal (LL) chemical group of ordinary chondrites, distinguished by their low siderophile element content, fairly large chondrules of $\approx$0.9 mm, and oxygen isotope compositions that are further above the terrestrial fractionation line than those of other ordinary chondrites\footnote{Taken from https://meteorites.asu.edu/chondrites/ll}), Young's modulus is 7.8$\times 10^{10}$ N m$^{-2}$, and the Poisson ratio is 0.25 \citep{schultz1995}.  Based on these parameters, the values of $k_n$ and $\gamma_n$ can be calculated as:
\begin{equation}
k_n=\frac{2Y\sqrt{r_{eff}}}{3(1-\nu^2)}
\end{equation}
\begin{equation}
\gamma_n=A_r\sqrt{\xi}
\end{equation}
where $r_{eff}$ is the effective radius of the colliding spheres of radii $r_1$ and $r_2$
\begin{equation}
\frac{1}{r_{eff}}=\frac{1}{r_1}+\frac{1}{r_2}
\end{equation}

The coefficient $A_r$ can be adjusted in order to manipulate the coefficient of restitution of the particles.  In our simulations, we have chosen to use two values: one while the particles are settling (5$\times 10^{-4}$) that provides a coefficient of restitution of 0.1 and another (2$\times 10^{-4}$) for the actual simulations which results in a coefficient of restitution of $\approx$0.5.  We have chosen this value as an average of what has been found experimentally for non-spherical particles \citep{wang2020}.  Having a very low coefficient of restitution during the settling period diminishes the settling time.

The particles are contained in a box with a solid bottom (15$\times$15 cm), horizontal periodic boundary conditions and a moving top that allows us to impose a very well determined confining pressure to the system.  The solid bottom is treated as a body with infinite  mass, whereas the moving top is treated as any other particle but with a single degree of freedom along the vertical axis.  The moving top has the same horizontal dimensions as the container, it is 4 cm thick and has a density of 3200 kg m$^{-3}$, exactly as the particles in the container.  There is nothing really special about the thickness or density of this body at this point; however, making it much less massive would make it much more sensitive to small forces (due to numerical inaccuracies) and the system would become less stable.  The settling procedure is as follows: initially, the particles are placed inside this box so that their centres form a hexagonal close packed lattice (HCP).  To produce an RCP arrangement, particle separation is 2.1 times the diameter of the largest particles so they have enough space to move around and avoid any initial overlap and they are given random speeds in all three axes of motion ([-0.15, 0.15]m/s).  After that, the code lets the particles settle under Earth's gravity.  Settling the particles only under the nominal confining pressure and no gravity or under lower levels of gravity would result in unnecessarily long settling times without any practical benefit.

Once that is done, the moving top is placed on the surface of the grains and the system is allowed to settle again.  Finally, the weight of the entire system is calculated, gravity is set to zero and a force equal to that weight is applied to the moving top so that the system is minimally disturbed and it is allowed to settle again.  Particle polydispersity and packing do have an influence over sound wave propagation \citep{ruiz2016} so in order to reduce the parameter space, all simulations are done with the exactly the same granular system up to this point.  After this, the force on the moving top is diminished by one order of magnitude and the system is allowed the settle for as many times as needed to reach the desired confining pressure.  During this process, the particles are also subjected to a Stokes' like drag \citep{sanchez, sanchez2014,zacny2018} so that the energy released when the pressure is reduced is quickly removed.  The aggregates so obtained had a filling fraction of $\approx$0.64.  We also considered the possibility to simply reduce the gravitational field as a means to reduce the pressure on the grains; however, this would have left the system with a pressure gradient that could have modified the results.  The time step ($\delta t$) that is used during the settling procedure is 2$\times 10^{-6}$ s for $P\leq2500$ Pa; this was changed to 1$\times 10^{-6}$ s if this condition was not met; not doing this results in artificially active systems as particle-particle overlap, which is directly related to the magnitude of the repulsive forces, would be excesive.  The experiment starts after all this lengthy settling procedure is finished and $\delta t$ was further reduced to 2$\times 10^{-7}$ s to allow for greater accuracy and consistency among all the simulations.  

In order to ensure that our systems are not artificially active, we ran simulations with several values of $\delta t$ and found that the values provided here were a good compromise between accuracy and speed.  The height of the settled system is approximately 1.77 m.  The accuracy of the confining pressure applied to the system is monitored at the bottom of the container and through the net force on the moving top; in general the variation from the nominal value applied to the moving top is $0.001\%-1\%$ for all the tested values.  Exactitude was increased as the confining pressure was increased.

The wave is initiated by the piston 0.5 s after the settling procedure has finished, the piston is given a vertical velocity (downwards in the simulation setup, see Fig.~\ref{sim-ini}).  This velocity will be either constant, in order to calculate the sound speed in the medium, or instantaneous (as a pulse), to study the attenuation of the system.  The entire system is divided into horizontal slices 5 cm thick so that more than a monolayer is observed at a time.  The kinetic energy (translational and rotational) of the particles in each slice is calculated in order to observe the wave transmission.  We define a wave in terms of the spatial distribution of the kinetic energy of the particles and its transmission in terms of the position of the wavefront (see Fig.~\ref{vel-prof-CVI0.1}).  Data is collected every 5$\times 10^{-5}$ s after the wave is started and this is done for 0.015 s which gives enough time for the wave to go through the system completely.  For the pulse experiments, we will be measuring the time that it takes for the peak of the wave to travel between the  centre of the 36$^{th}$ slice and the centre of the 1$^{st}$ slice from the bottom (1.775 m) as there is only one pulse that will be generated in any simulation.  Also, the confining pressure will be set at 0.1, 1, 10, 100, 1000, 10000 and 50000 Pa; simulations will run for 0.6 s in total.  Greater confining pressures have already being explored in the available literature and are out the scope of this paper.

In order to measure the effect of the passing wave on the granular medium, we also calculate the stress tensor and the principal stresses of the particles in each slice.  As in our previous research \citep{sanchez2012}, this is done by calculating the average Cauchy stress tensor.  This is, the stress tensor of an aggregate of particles contained within a volume {\it V} is equal to the sum of the dyadic product of the reaction force ($\vec f$~) between any two particles in contact and the arm vector ($\vec l$~) between their centroids  \citep{janthony2004}.  Mathematically:
\begin{equation}
\bar\sigma=\frac{w}{V}\sum_{c\in K_V}\vec f \otimes \vec l
\end{equation}

In the above equation $K_V$ is the set of contacts between the particles in the volume $V$ and the weight function $w$ has a value of 1 for contacts of particles in the same volume and 0.5 when they are in different volumes \citep{luding2008}.  By doing this the contribution to the stress provided by two particles which centres are contained in the same volume will be fully added to that volume, whereas if the particles are contained in different volumes, the stress will be equally divided between them.   Once the stress tensor is calculated, we calculate their eigenvalues and from them, the hydrostatic pressure.  For this specific simulation, we will calculate the stress tensor over each slice of the container and over the entire container.  The former will provide a very accurate description of the stress state of the system as the wave progresses whereas the latter will provide a space-averaged description of the medium.

\section{Results}
\subsection{Constant Velocity Compression}

\begin{figure}[h!]
\begin{center}
\includegraphics[scale=0.9]{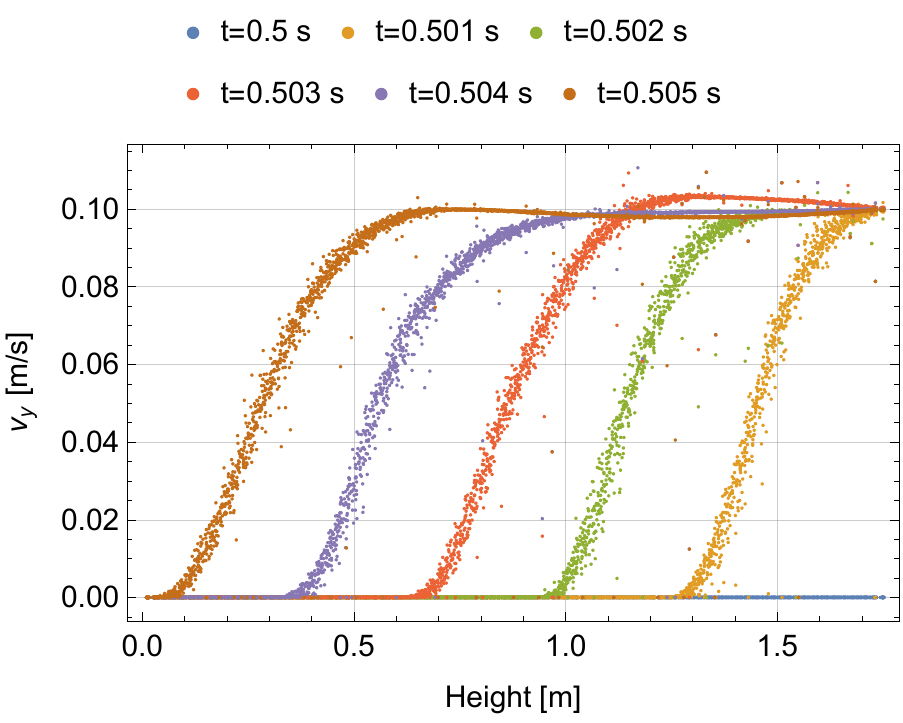}
\caption{Time evolution of the  vertical velocity profile of the particles inside the container at 0.1 Pa.  Piston speed set at 0.1 m/s.}
\label{vel-prof-CVI0.1}
\end{center}
\end{figure}

\begin{figure}[h!]
\begin{center}
\includegraphics[scale=0.9]{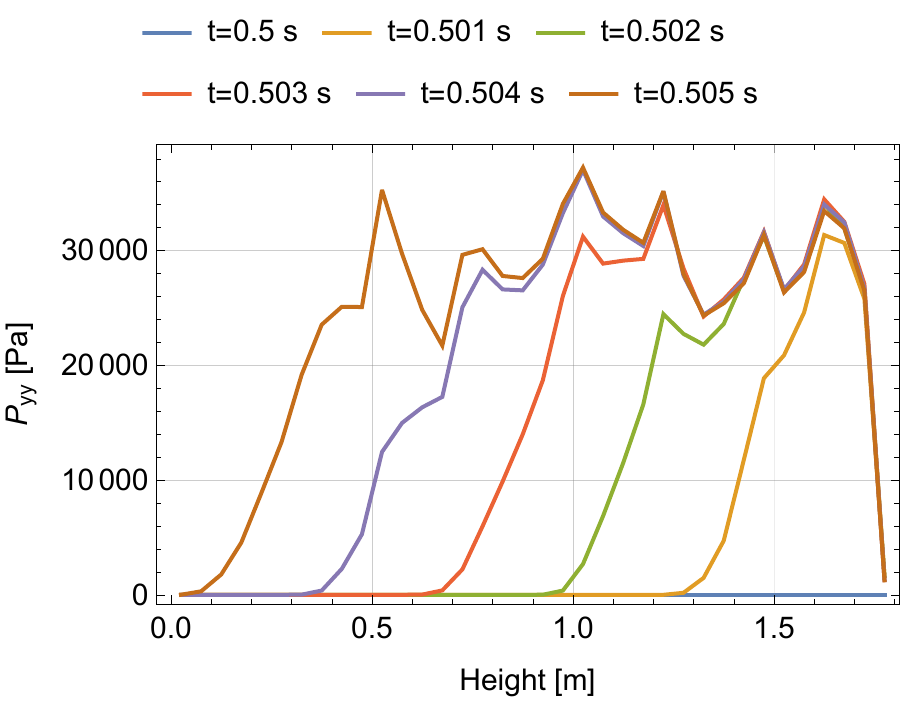}
\caption{Time evolution of the $P_{yy}$ component of the stress tensor of the particles in each slide inside the container at 0.1 Pa.  Piston speed set at 0.1 m/s.}
\label{hpress-C0-P0.1-CVI01}
\end{center}
\end{figure}

In order to measure the speed of sound within the simulated granular media, we simulate a piston-compression experiment.  As explained above, the grains settled under different confining pressures generated by a moving piston.  Now we are going to use this piston, compressing the medium at a constant speed, to generate a wavefront.  The advancement of this pressure wavefront will allow us to measure the sound speed in the medium \citep{gomez2012}.  The compression speeds we tested are 0.001, 0.005, 0.01, 0.1, 0.5, 2, 3.5 and 5 m/s, the particles are always cohesionless and the tested confining pressures are 0.1, 1, 10, 100, 1000, 10000 and 50000 Pa. 

Fig.~\ref{vel-prof-CVI0.1} shows the time evolution of the vertical velocities ($v_{y}$) of all the particles for a confining pressure of 0.1 Pa and a compression speed of 0.1 m/s.  Fig.~\ref{hpress-C0-P0.1-CVI01} on the other hand show the corresponding evolution of the $P_{yy}$ component of the stress tensor inside each individual slice of the granular medium.  Both figures show the advancement of the wavefront at what appears to be a constant rate.  Measuring this rate of advancement is what allows us to calculate the speed of the seismic wave.

\begin{figure}[h]
\begin{center}
\includegraphics[scale=1.1]{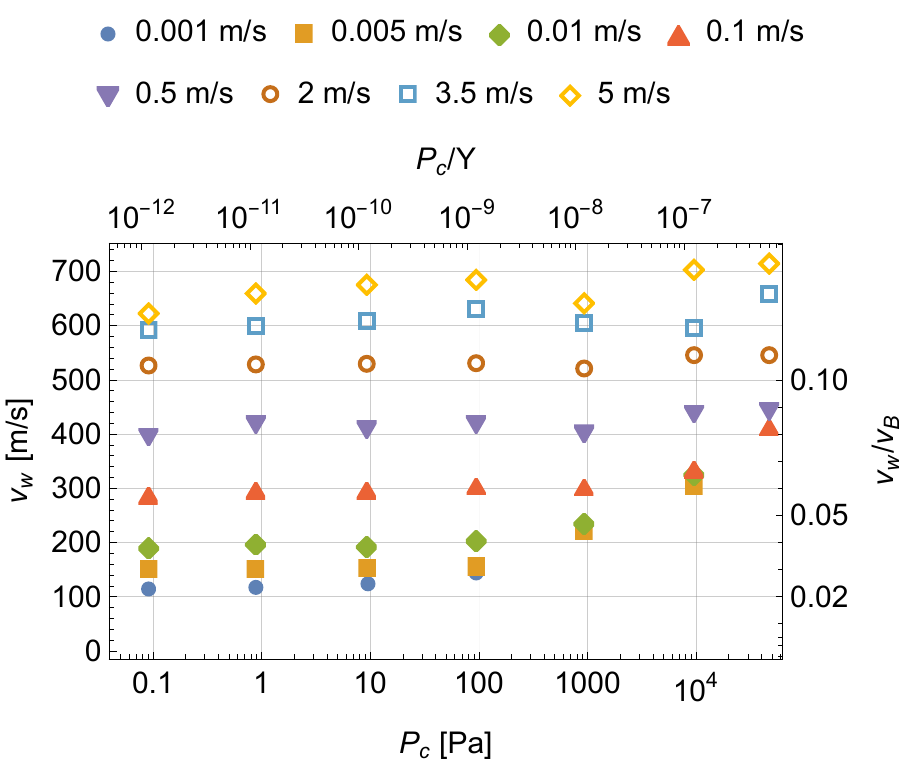}
\caption{Wave speed vs.~confining pressure in the simulated systems.  The different colours represent the different compression speeds of the piston.  The horizontal axis above has been scaled by the Young's modulus of the material, whereas the vertical axis at the right has been scaled by the wave velocity in basalt, $v_B=5\times10^3 m/s$.}
\label{PvsVw}
\end{center}
\end{figure}

In Fig.~\ref{PvsVw} we have plotted the measured wave speed as a function of the confining pressure ($P_c$) of the different systems; showing that different wave speeds can be obtained at the same confining pressures.  In this plot, different symbols represent data points that correspond to the different compression speeds that were used; furthermore, it seems there is a minimum wave speed for each confining pressure.  Additionally, data points that correspond to the same compression speed are almost in a horizontal line as the value of $v_c$ increases, which means that $P_c$ is important at very low compression speeds, but compression speeds is more important as it increases.

In order to understand this behaviour, we observe in Fig.~\ref{hpress-C0-P0.1-CVI01} that, as time passes, the pressure in the compressed section of the medium begins to stabilise.  Once the wavefront has passed, $P_{yy}$ stops changing.  This  is observed in all our simulations of this type.  This being so, it is possible to define what we call a ``stable pressure'' ($P_s$) as an average of the $P_{yy}$ component of the stress tensor of each slice in which the pressure has stabilised.

\begin{figure}[h]
\begin{center}
\includegraphics{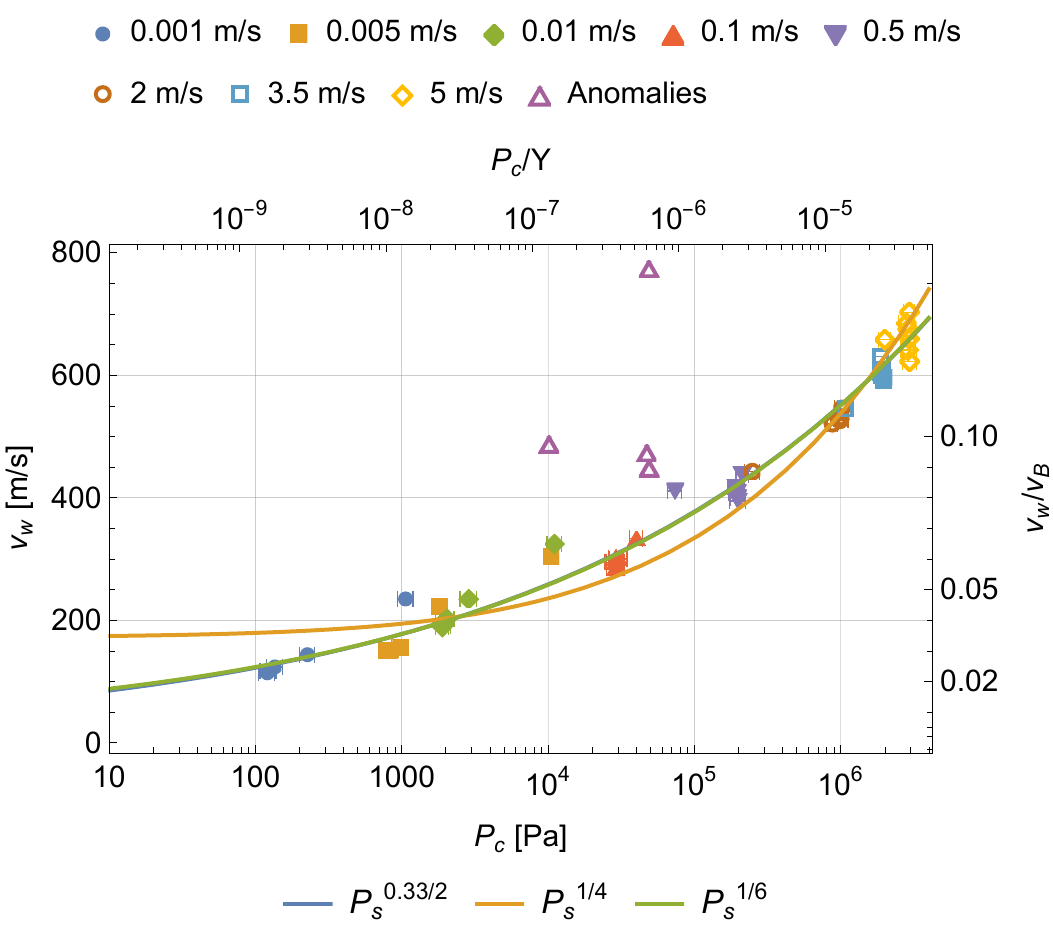}
\caption{Seismic wave speed vs.~stable pressure for all the constant velocity compression simulations  The three trend lines have been added to show how different pressure exponents adjust the data points.  The data points represented by the purple triangles ({\color{purple}$\bf\triangle$}) labeled as {\it Anomalies} were not taken into account.  The horizontal axis above has been scaled by the Young's modulus of the material, whereas the vertical axis at the right has been scaled by the wave velocity in basalt, $v_B=5\times10^3 m/s$.}
\label{CVI-speed}
\end{center}
\end{figure}

In Fig.~\ref{CVI-speed} we plot the seismic wave speed as a function of the stable pressure ($P_s$), and three trend lines that fit the data points to the expression
\begin{equation}
v_w=\sqrt{a+bP^c}.
\end{equation}

We add the trend lines as the main prediction of the theory \citep{goddard1990,makse1999} is that the scaling of the bulk modulus $K$ and shear modulus $\mu$ vary with the pressure as $P^{1/6}$.  So for vanishingly small confining pressures, such as those existing in the asteroid environments, $v_w$ should tend to zero.  However, there is experimental evidence showing that for small confining pressures, the dependency changes to $P^{1/4}$ \citep{goddard1990}.  Table \ref{Wspeed-table} collects the values of $a$, $b$ and $c$ for the three different trend lines in Fig.~\ref{CVI-speed}.

\begin{table}[]
\caption{Fitting parameters for the trend lines in Fig.~\ref{CVI-speed}.}
\begin{center}
\begin{tabular}{c c c c c}
\hline
Trend Line& $a$	&	$b$	&	$c$&		$v_w$ at $P=0$\\
\hline
\hline
{\color{f1}\rule{1cm}{2pt}}	&481.663	&3159.51	&0.330296&21.9\\
{\color{f2}\rule{1cm}{2pt}}	&29533.8	&259.201	&1/2&171.9\\
{\color{f3}\rule{1cm}{2pt}}	&1240.03	&3020.76	&1/3& 35.2\\
\hline
\end{tabular}
\end{center}
\label{Wspeed-table}
\end{table}

Note that for the trend lines with $P^{1/4}$ and $P^{1/6}$, the value of $c$ was fixed, whereas for $P^{0.33/2}$ the exponent was also one of the fitted parameters.  However, as it can be observed the trend lines resulting from $P^{1/6}$ and $P^{0.33/2}$  are almost indistinguishable from one another.  This result agrees with the theory as long as the pressure that is accounted for is what we have denominated ``stable pressure'' and not only the confining pressure initially provided by the moving piston.  Also, we are not observing a $P^{1/4}$ dependency as that is a purely experimental result, not yet fully explained \citep{goddard1990, makse1999,velicky2002}, and our simulated system is as ideal as the theory requires; this is, cohesionless, perfectly spherical grains under a constant confining pressure.

Though fig.~\ref{CVI-speed} is similar to fig.~3 in \cite{gomez2012}, their figure shows a relationship between wave speed and compression speed and ours is between wave speed and confining pressure.  Additionally, the theory they developed is for a 2D system and so their applicability to a 3D system, like ours, cannot be taken for granted.  In fact, if we use their eq. 2 to calculate the wave speed, the results are consistently $\approx$3.2 times higher than our measured values.

The data points termed as {\it Anomalies} where not taken into account for the calculation of any of the trend lines.  These data points were observed for $P$=10 kPa at 0.001 m/s compression speed, and $P$=50 kPa at 0.001, 0.005 and 0.01 m/s compression speed.  This is, the highest simulated pressures and the slowest compression speeds.  This could be due to numerical inaccuracies that are more evident for these specific cases and could possibly be resolved by reducing even more our $\delta t$ for the integration of the equations for motion.  However, solving this problem is out of the scope of this paper.

\begin{figure}[h]
\begin{center}
\includegraphics[]{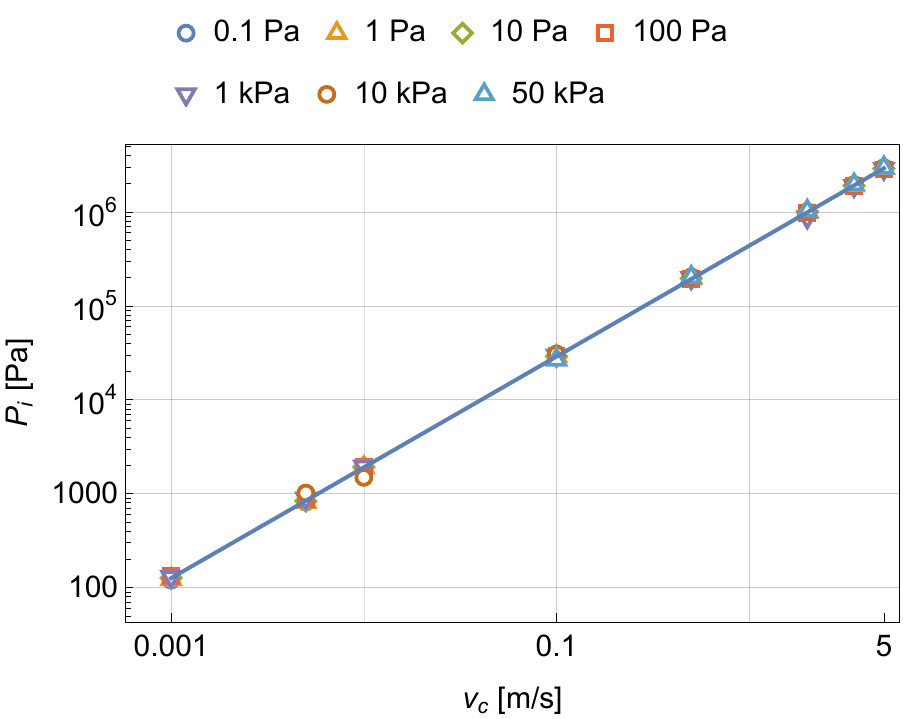}
\caption{Compression speed vs.~the pressure induced by the compression.  The different symbols correspond to confining pressure of the system in which the measurement was carried out.  The blue line ({\color{f1}\rule{0.5cm}{2pt}}) is a fit of the data with the equation $P_i=4.36\times10^5 v_c^{1.18}$.}
\label{vcvsPi}
\end{center}
\end{figure}

This far we have shown that wave speed is related to $P_s$, and we have also seen that the value of $P_s$ changes depending on $v_c$ even for systems at the same confining pressure.  This tells us that the compression driven by the motion of the piston is adding to the confining pressure, that this addition results in the observed stable pressure and that this added/induced pressure ($P_i$) is directly related to the compression speed.  If the pressure that is induced by the motion of the piston is independent of the value of $P_c$  (at least within the low values tested here) a plot of the compression speed versus the induced pressure should show a straight line and several data points on the same position corresponding to simulations with the same compression speed.  Such a plot can be observed in fig.~\ref{vcvsPi}, where all these characteristics are present.  The value of the induced pressure has been calculated as $P_i=P_s-P_c$.

Fig.~\ref{vcvsPi} confirms that there is a very clear relationship between $P_i$ and $v_c$, shows that the former can vary from $\approx 10^2 - 10^6$ Pa, and establishes that $P_s$ can be defined as $P_i+P_c$.  This would be the reason why as $v_c$ increases so does $v_w$.  If $Pc\gg P_i$, as for the cases with very small $v_c$ and $P_c$ in the kPa regime, $P_c$ will mostly determine wave speed.  On the other hand, if $Pi\gg P_c$, as for compression speeds in the m/s regime, $P_i$ will mostly determine wave speed.  As $v_c$ increases, $P_i$ increases and this causes $v_w$ to increase.  As an example, the smallest $v_c$ was 0.001 m/s and for these cases $P_i\approx 120$ Pa ,so regardless how small the value of $P_c$ could be, $v_w$ has to conform to that value of $P_i$.  This is why in Fig.~\ref{PvsVw}, all data points at the same $v_c$ show a plateau at small confining pressures, but the plateau disappears at high compression speeds.

\subsection{Transmission of a Pulse}
The average impact velocity among colliding asteroids in the Solar System, as well as the most probable one, is $\approx$5 km/s range \citep{bottke1994}; this puts these collisions in the hypersonic regime.  The simulations presented in this paper are ill suited to address how this collision will initially evolve as there will be chemical and physical processes that have not being coded in the simulations, comminution, heat transfer or even melting.  However, at some point, within the interior of the asteroid, the shock wave produced by the impact will lose enough energy to become a seismic wave and it is here were our results can potentially be applied.  

Measuring the maximum amount of energy per unit area (the peak) that can be transmitted through the asteroid will bound the size of the particles that could be lofted and their speeds.  This lofting could potentially trigger particle flows, resurfacing, particle ejection, crater erasure or particle size segregation \citep{wright2020}.  In the particular case of the impacts generated by the Haybusa2 and DART missions, in principle, the same should be expected.  The impact on asteroid Ryugu used a 2 kg projectile at $\approx$2 km/s \citep{arakawa2017}, the impact on asteroid Dimorphos will use the entire mass of the spacecraft (610 kg) at an impact velocity of $\approx$6.5 km/s \citep{cheng2020}.

For these simulations, the piston is given an instantaneous vertical velocity downwards and is free to move vertically.  We do this to study how the energy of a low velocity impact is transmitted and dissipated in a granular medium that simulates the asteroid environment.  The impact speeds will be 0.01, 0.1, 0.5 and 5 m/s, and the confining pressure, 0.1, 1, 10, 100, 1000, 10000 and 50000 Pa.

\begin{figure}[h]
\begin{center}
(a)\includegraphics[scale=0.9]{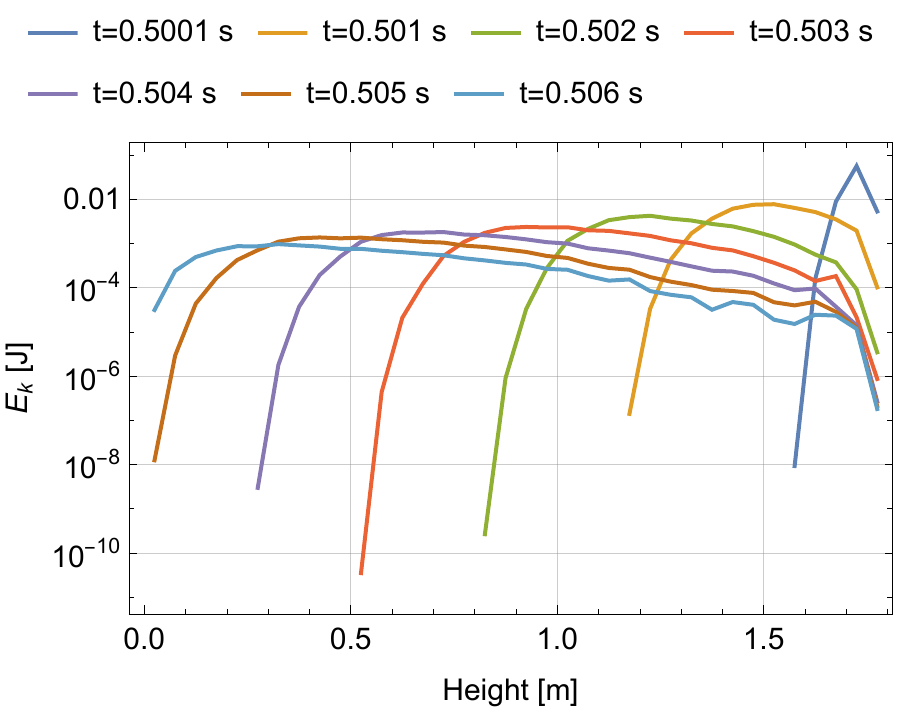}
(b)\includegraphics[scale=0.9]{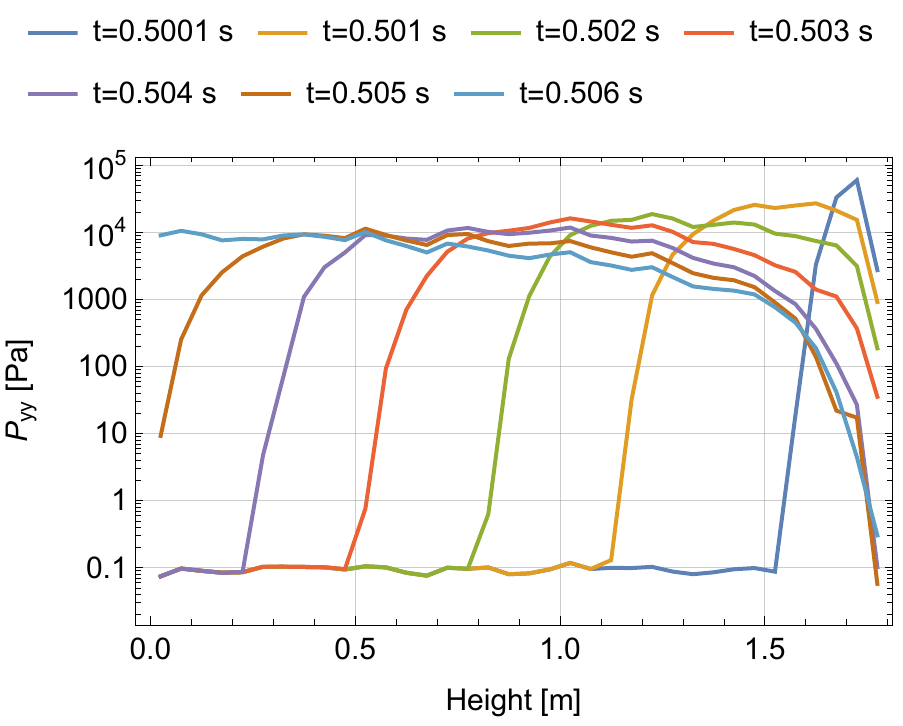}
(c)\includegraphics[scale=1]{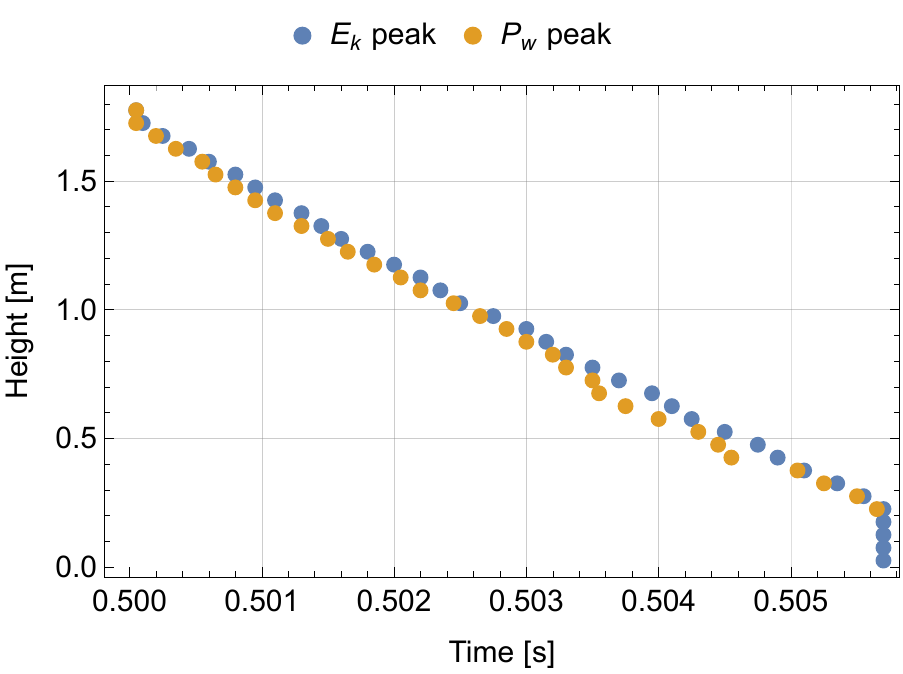}
\caption{a) The kinetic energy of the particles vs. the height and over time, b) the $P_{yy}$ component of the stress tensor vs. the height and over time and, c) the position of the peak of both, kinetic energy and $P_{yy}$ vs. time.}
\label{Press-KE-Peaks}
\end{center}
\end{figure}

The impact is produced at t=0.5 s and data is collected every 0.00005 s for 0.03 s which provides enough time for the peak in kinetic energy to travel through the entire container.  The peak is is said to be initiated its travel at t=0.5 s and it is said to have arrived to the last slice at the bottom of the container at the instant at which the kinetic energy of that slice has reached a peak.  Figs.~\ref{Press-KE-Peaks} shows, for a cohesionless ($\sigma_{yy}$=0) system, at an confining pressure of 0.1 Pa, that is impacted by a piston at 0.5 m/s: a) the kinetic energy of the particles vs.~the height at intervals of 0.001 s after the impact b) the $P_{yy}$ component of the stress tensor vs.~the height at intervals of 0.001 s after the impact, c) the position of the peak of both, kinetic energy and $P_{yy}$ vs.~time. These measurements allow us to calculate the velocity of transmission of the peak of the kinetic energy produced by the impact on the system.  They also show how energy is quickly dissipated as the peak reaches the bottom of the container.  Additionally, fig.~\ref{Press-KE-Peaks}c shows little variation in the transmission speed of the peak, be this the kinetic energy or pressure, across the length of the granular system.  This means that measuring the velocity of transmission of the peak in kinetic energy or pressure produced by the pulse, the results should not vary by much even if, as can be observed, the peak in pressure tends to precede the arrival of the peak in kinetic energy.  Figs.~\ref{Press-KE-Peaks} (a) and (c) show that as time passes, there is  a decay of the pulse amplitude, and a broadening of the pulse, similar to what has been observed at higher confining pressures \citep{langlois2015,quillen2022}.  Pulse broadening  arises due to variations and irregularities in the force contact network: variations in travel times along different force contact chains \citep{owens2011},  irregularities in the force contact network \citep{hostler2005}.  Energy can be dissipated (leading to attenuation) via several mechanisms, including frictional and inelastic particle interactions (discussed by \citealt{hostler2005}), particle rearrangements (discussed by \citealt{zhai2020}),  scattering through the particle contact network \citep{owens2011} and scattering due to variations in particle stiffness \citep{langlois2015}, variations in the packing fraction or porosity and variations in the connectivity of the particle contact network.

\begin{figure}[h]
\begin{center}
\includegraphics[scale=1]{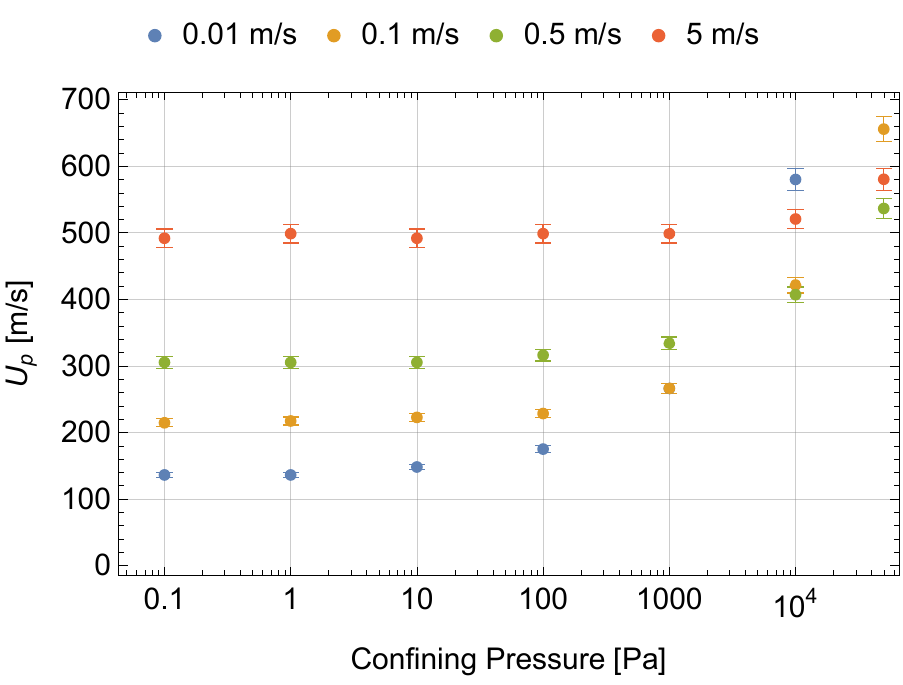}
\caption{Velocity of transmission of the peak ($U_p$) of the energy of the impact through the granular medium versus the confining pressure applied to the system.}
\label{OP-vs-Up}
\end{center}
\end{figure}

Fig.~\ref{OP-vs-Up} shows the velocity of transmission of the peak of the energy of the impact through the granular medium versus the confining pressure applied to the system.  The plot shows the formation of plateaus at low enough pressures which initiation depends on the speed of the impact.  These are similar to those observed by \cite{wildenberg2013} but in this case they are measuring the wave speed and not only the speed of transmission of the peak in kinetic energy.  More significantly, this structure is the same as show in Fig.~\ref{PvsVw}; which would imply a similar reason.  This is, that the pressure that is induced by the pulse is also controlling the transmission speed of the pulse.  However, given the decay in amplitude and the broadening of the pulse, defining a stable pressure, as we did for the compression simulations, is impossible.

\begin{figure}[htbp]
\begin{center}
\includegraphics[scale=1]{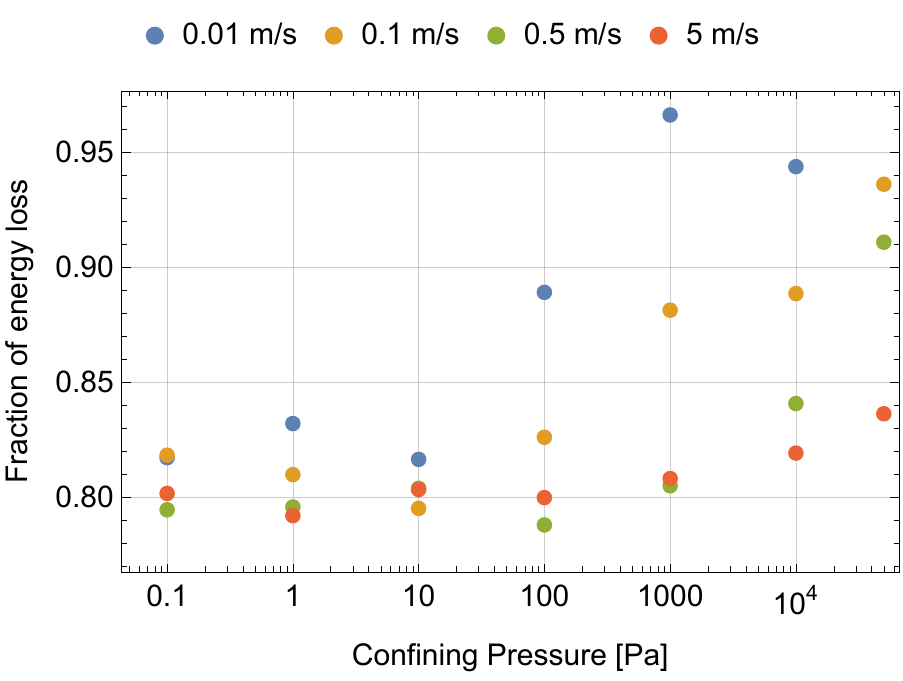}
\caption{Fraction of energy loss for all the values of confining pressure and impact speed not taking into account the initial energy of the impactor.}
\label{EnergyLoss}
\end{center}
\end{figure}

The measurement of $U_p$ provides us with a way to know the appropriate time at which energy dissipation should be measured given that the experimental setup is finite.  Fig.~\ref{EnergyLoss} shows the fraction of the initial energy of the impactor that had already been lost by the system at the time the peak of the kinetic energy of the particles arrived at the bottom of the container.  Notice that we are calculating this loss as the difference between the maximum amount of energy that was transmitted to the grains due to the impact and the amount that the grains have at the moment the peak of the wave arrives at the bottom of the container; we are not using the energy of the impactor as a point of comparison.  For the investigated range of parameters, this fraction varies between $\approx$0.8-0.95.  If the initial energy of the impactor is considered instead, the fraction of energy that is lost goes from $\approx$0.95-0.99.  This finding agrees with other sets of simulations \citep{sanchez-lpsc2020,sanchez-lpsc2021,sanchez-pg2021} carried out with in a shorter container, but with the same material parameters.  We will use this later estimate of energy loss in the next sections given that usually what is known is the energy of the spacecraft at the moment of contact.  As it can be noticed, this figure shows that energy dissipation increases as the confining pressure increases; in particular, there is a peak for $P_c=1 kPa$.  As explained by \cite{fonseka2022}, energy dissipation increases with confining pressure in the weak shock regime as strongly pre-loaded contacts dissipate more energy.  They obtained this result for simulated constant velocity compression experiments, similar to ours and with a similar simulation code, but in 2D.  Their lowest confining pressure was higher than ours, but some of the compression velocities are comparable.  We are applying their results to our pulse transmission experiments because within the simulation method used, energy dissipation between two particles in contact is independent of the reason for the relative movement of the grains.

\begin{figure}[]
\begin{center}
(a)\includegraphics[scale=0.9]{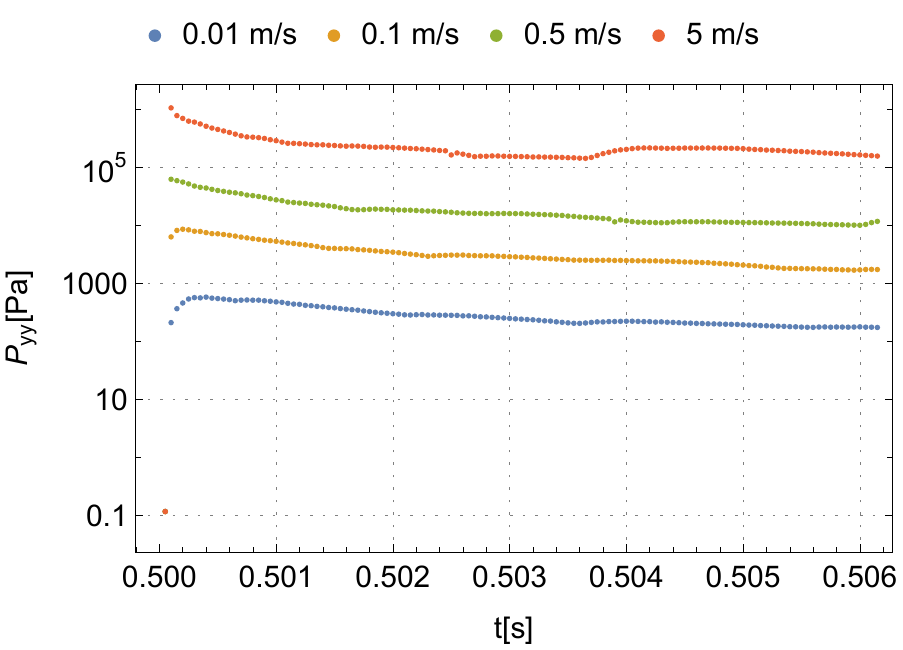}
(b)\includegraphics[scale=0.9]{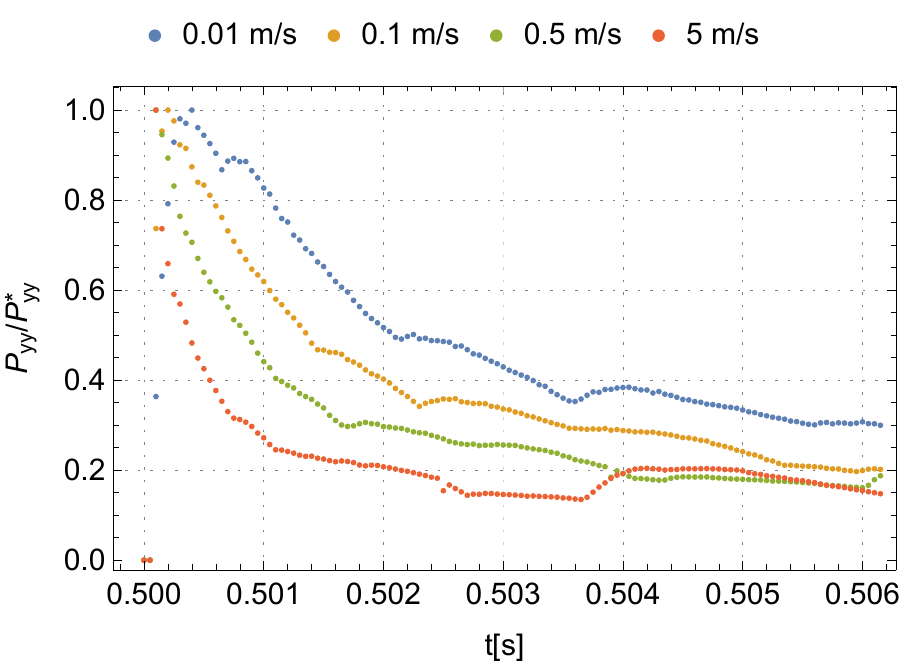}
(c)\includegraphics[scale=0.9]{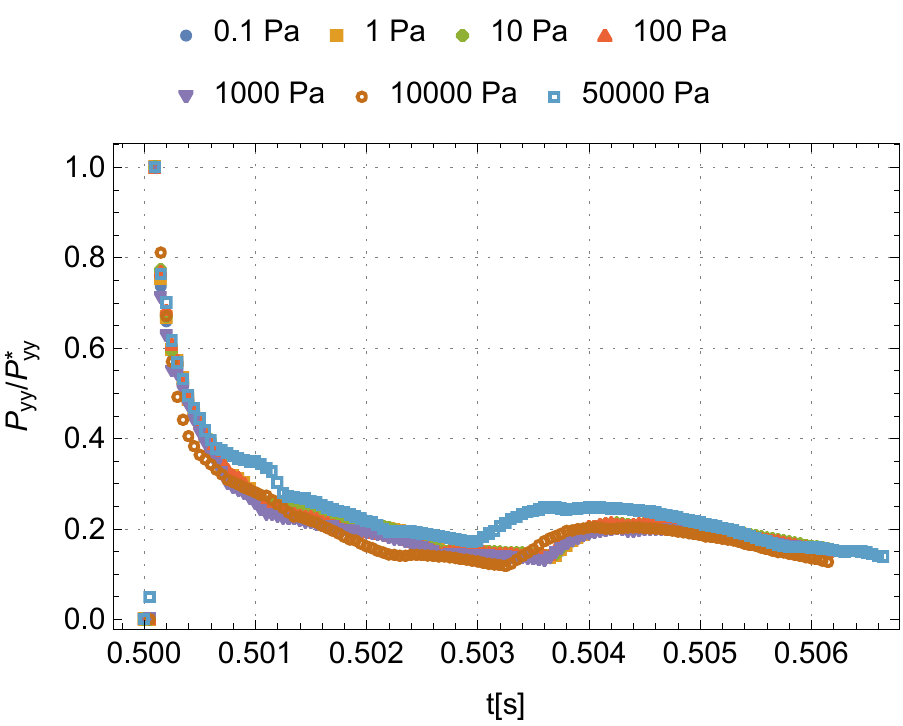}
\caption{Amplitude of the pressure peak generated by different impacts over time as it is transmitted through the granular medium. (a) Confining pressure of 0.1 Pa, (b) normalised amplitude of the same system (confining pressure 0.1 Pa, $P_{yy}^*$ is the peak value of the pressure generated by the pulse), (c) normalised amplitude for 5 m/s impacts at different confining pressures.}
\label{pulse-attenuation}
\end{center}
\end{figure}

Fig.~\ref{pulse-attenuation} shows the attenuation of the peak of the pressure pulse as it moves through the granular media.  Fig.~\ref{pulse-attenuation}(a) shows the attenuation for a confining pressure of 0.1 Pa and various impact speeds.  Fig.~\ref{pulse-attenuation}(b) shows the attenuation of the same pressure pulses, but normalised with respect to the value of their individual initial peaks.  It is evident that the rate of attenuation of the peak pressure increases with the impact speed.  Similar plots can be obtained for confining pressures of up to 50 kPa. Fig.~\ref{pulse-attenuation}(c) shows the attenuation of the peak of the pulse for a single impact speed (5 m/s) and all the tested confining pressures.  As it is evident, all the points collapse in a single curve up to the instant when the peak of the wave reaches the bottom of the container.  This however, does not happen for all impact speeds; in fact, the curves tend to collapse as the impact speed increases.  This suggests that the rate of attenuation of the peak of the wave is independent of the confining pressure only for high enough pressures and so it is the total pressure, this is the sum of the confining pressure and the pressure induced by the pulse that controls the attenuation of the peak.

\subsection{Impactor Energy and Momentum}

This far we have explored how the energy of the impact is transmitted through the granular medium.  However, we have said nothing about the impactor itself and it is unclear whether it is its energy or momentum what sets the value of $U_p$.  Understanding this would provide us a way to scale and extrapolate our results to impact regimes in which DEM simulations would be too time consuming as the momentum would be too high to make simulations practical.  Note that this extrapolation should not reach the hypervelocity regime as at those impact speeds other chemical and physical phenomena, not coded in our simulations, also take place.

In order to do this, the simulation will have to change a little. We will initiate the wave with the piston; we will change its thickness so changing its mass, we will also change its impact velocity.  As before, the piston will be provided with an instantaneous vertical velocity at t=0.5 s; everything else will be the same.  The thickness of the piston was set to: 4, 8 and 16 cm; the impact speed was: 0.1, 0.2, 0.4, 0.8 and 1.6 m/s.  The nominal confining pressure was kept at 10 Pa to minimise the settling time and to keep the accuracy of the numerical method.

\begin{figure}[htbp]
\begin{center}
\includegraphics[scale=0.9]{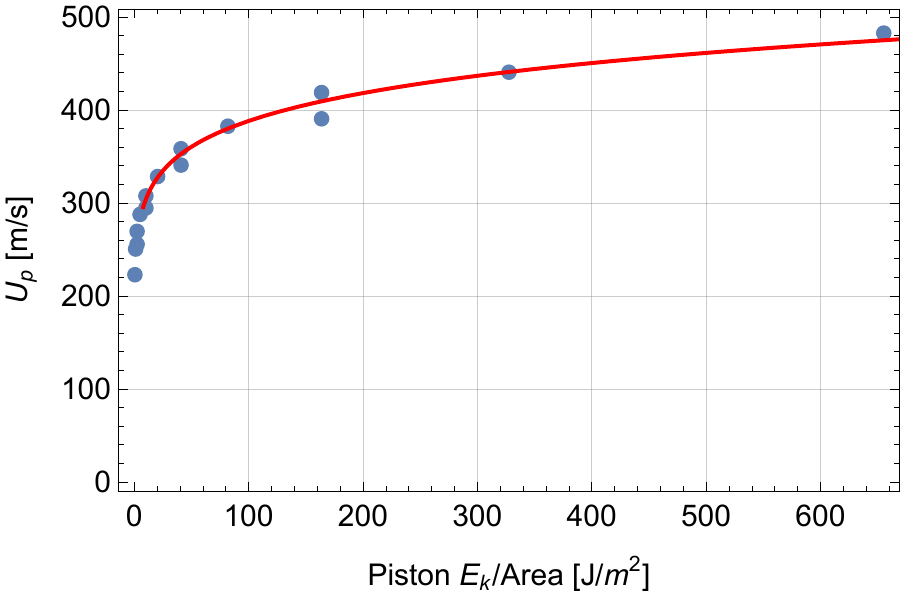}
\includegraphics[scale=0.9]{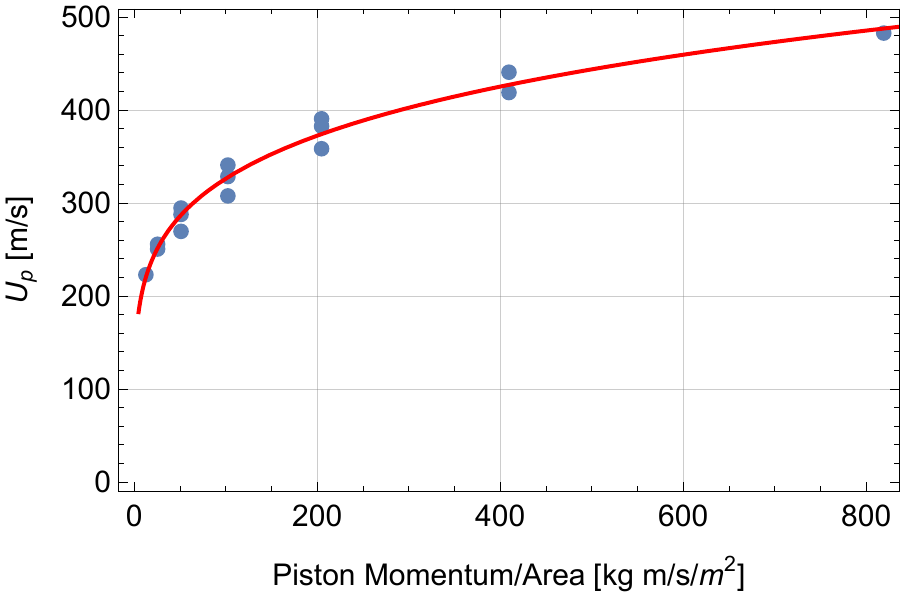}
\caption{(Left)Kinetic energy of the piston per unit area vs. the speed of the peak in kinetic energy of the system.  (Right) Momentum of the piston per unit area vs. the speed of the peak in kinetic energy of the system.}
\label{scaling}
\end{center}
\end{figure}

We have designed the simulations in such a way that, by systematically varying impact speed and piston height, a few of them would have exactly the same values of either energy per surface area or momentum per surface area for the impact of the piston.  Figs.~\ref{scaling} show the results of these simulations in which we have also added trend lines.  From them it is difficult to choose one scaling over the other within the explored parameter space.  The two of them can be approximated by a power law and similar values of kinetic energy (or momentum) will produce similar transmission velocities for the peak.  The equations for the trend lines are:
\begin{eqnarray}
U_p&=&236.7 E_{kp}^{*0.11}\\
U_p&=&135.1 M_{p}^{*0.19},
\end{eqnarray}
where $E_{kp}^{*}$ is the kinetic energy and $M_{p}^{*}$ is the momentum of the piston, both per unit area and at the moment of impact.

\section{Asteroids and Impacts}

At the time of the writing of this paper, there have been at least three space missions to small asteroids, JAXA Hayabusa \citep{fujiwara} and Hayabusa2 \citep{watanabe2017} missions to asteroids Itokawa and Ryugu and the NASA OSIRIS-REx \citep{lauretta2012osiris} mission to asteroid Bennu (currently on its way back to Earth).  Apart from these, there is the (now defunct) NASA ARRM mission, the NASA DART mission to binary asteroid Didymos and the ESA HERA mission to the same asteroid.  The last one has not launched yet.  Additionally, the ESA ROSETTA mission successfully landed a pod on comet Churyumov-Gerasimenko 67/P \citep{biele2015}.  All these missions have one thing in common; they all had, or will have, interactions with the surfaces of their target bodies and as far as we have observed and calculated, these surfaces are granular and cohesive though cohesive strength is very weak compared to terrestrial standards.  Additionally, asteroid evolution is collisional, but the frequency of the impacts depend on the size and velocity of the impactor and the target \citep{bottke1994}.  Given the low escape velocity on small bodies, the landing and sampling operations are carried out in the centimetre per second range; however, explosions and impacts  are much more energetic and so the question of how the surrounding terrain, or the asteroid as a whole is going to be perturbed has to be taken into account.  This all is then strictly related to how energy is transmitted through and absorbed by  the impacted body which is, at a very basic level, a granular medium at very low pressure.  Even if the impacts are at hypervelocity, at some point, the impact waves have to be transformed into seismic waves.

\subsection{Low Velocity Impacts}

Examples of this kind of impact are those generated by the landing of the Hayabusa mission on asteroid Itokawa, the landing of the ROSETTA mission on comet Churyumov-Gerasimenko 67/P and the Touch-and-Go (TAG) experiment of the OSIRIS-REx mission.  Additionally, the impact on asteroid Didymos of the debris generated by the impact of the DART spacecraft will also be in this regime.  All these impacts take place in the m/s regime and so our results should be applicable.

All the simulations carried out by us and presented in this and previous papers \citep{sanchez-lpsc2020,sanchez-lpsc2021} have shown that energy dissipation is almost a constant across most values of the studied parameters, so we will use this to make some comparisons.  This is, about 95\% of the energy is dissipated by the time the peak of the wave has arrived to the end of the container.  Energy transmission per unit area then would go as $0.05^{R/1.77}$, where $R$ is the distance from the impact point if the surface area does not change.  So in general, energy per unit area ($E_{kf}^*$) would decay roughly as:
\begin{equation}
E_{kf}^*=\frac{E_{k0}}{A_0} T^{R/\lambda},
\label{E_decay}
\end{equation}
where $E_{k0}$ is the kinetic energy of the impactor at the moment of impact, $T$ is the transmission rate (0.05 in this case as there is a 95\% loss), $\lambda$ is a characteristic distance traveled by the peak of the kinetic energy so that it decays by 95\% (in this case, 1.77 m), $A_0$ is the area of contact of the impactor and $R$ is the distance from the impactor.  If we add to this the 3D nature of the system, energy per unit surface area would go as $R^{-2}$.  We realise that the effective area of the hemisphere over which the energy is transmitted should be reduced by $\phi$, the filling fraction of the granular medium.  However, $\lambda$ already contains that implicit dependence as its value was obtained from our simulations and the medium is porous by construction.  

\begin{figure}[h]
\begin{center}
\includegraphics[scale=0.5]{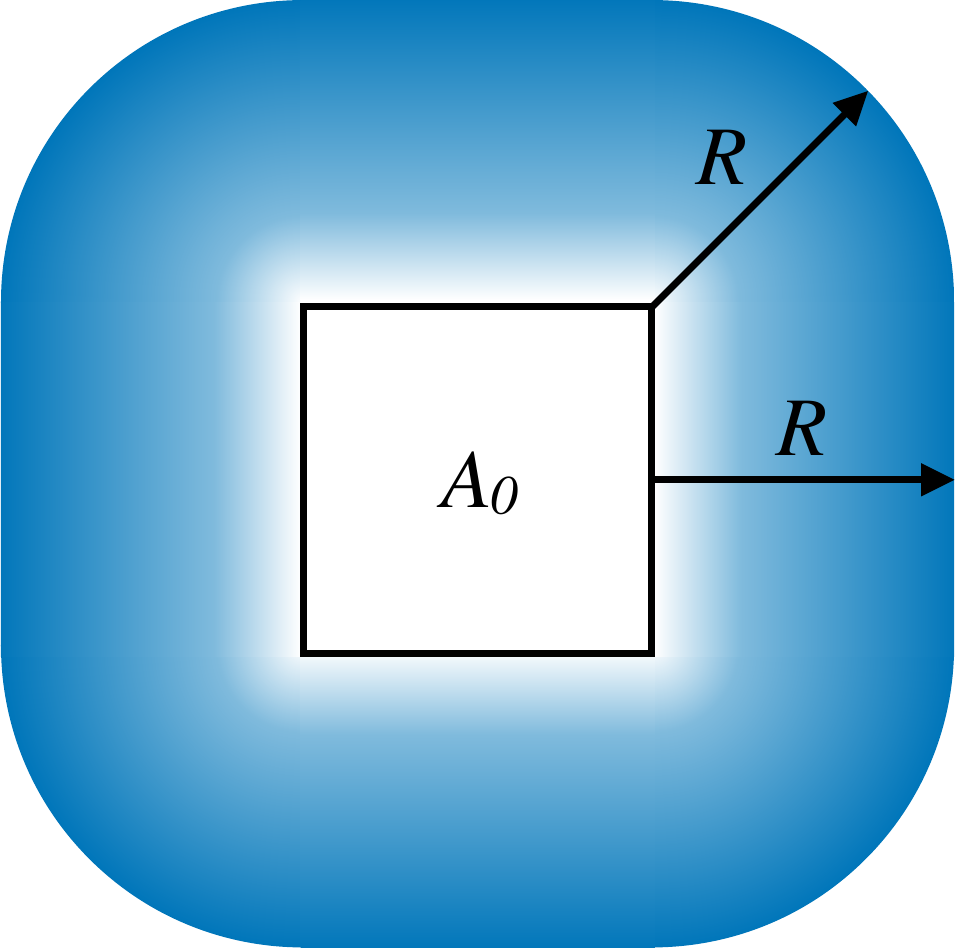}
\includegraphics[scale=0.5]{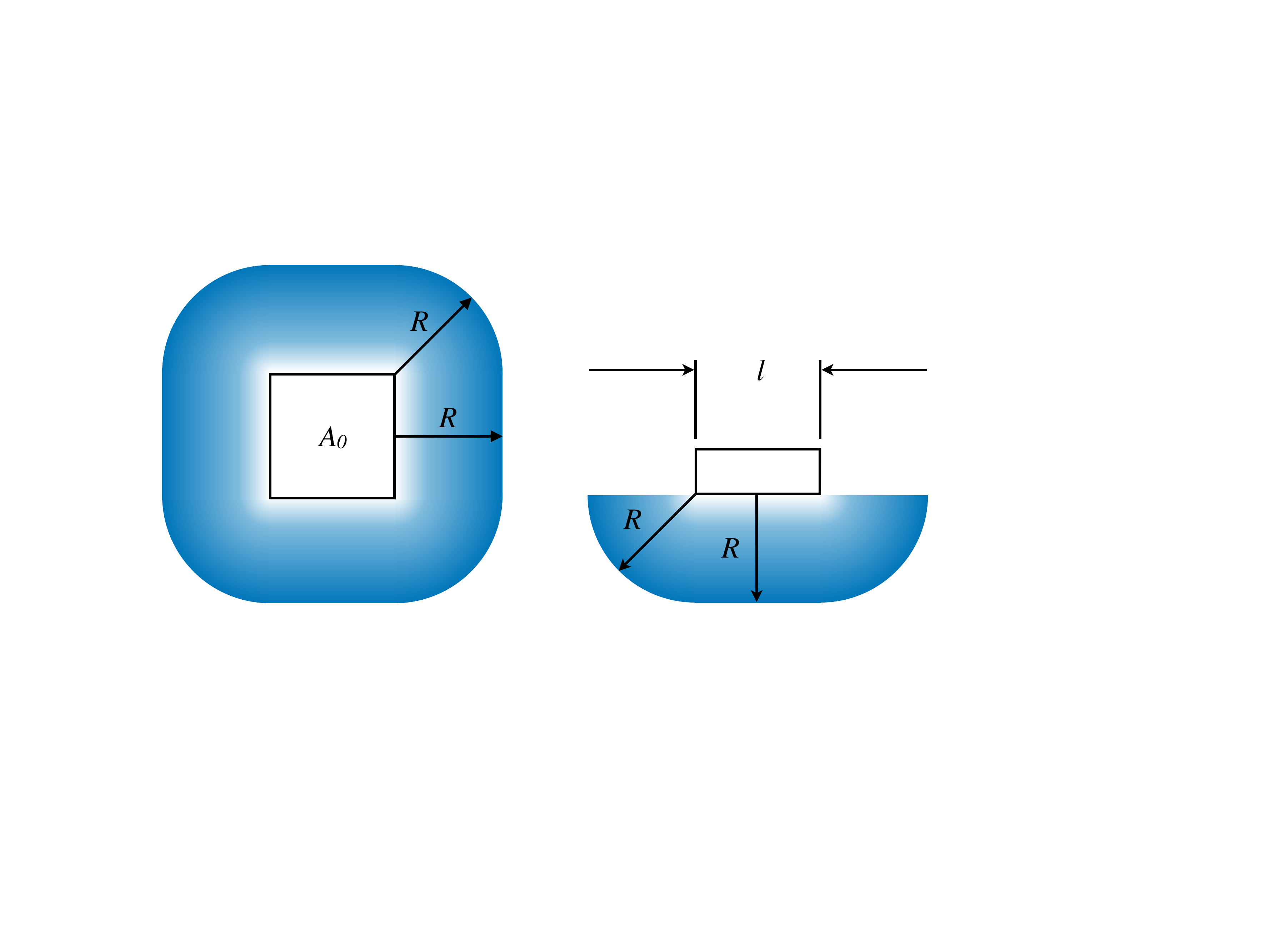}

\caption{Schematic view of the assumed expansion of the wave inside a 3D body.  $A_0$ is the surface area in contact with the granular medium, $R$ is the distance from any point on the impactor to the closes point in the wavefront, and $l$ is the side of the body if the shape of $A_0$ where a square.}
\label{wave-drawing}
\end{center}
\end{figure}

In order to calculate how the energy of the impact of an extended object is going to be spread inside a 3D body we will have to make some assumptions and simplifications.  We will start by assuming that the expansion is going to be spherical from every point on the surface of the impactor, that the impactor is square and that it makes complete contact with a flat surface.  Note that surface waves do not expand at the same speed as body waves and we are disregarding this fact.  However, as far as we are aware, no study has been carried out to understand how these are behave at extremely low confining pressures have been carried out (including ours).  So for the time being, we will assume that there is no difference between body waves and surface waves. Then, we can say that from each corner, the wave expands as an of a sphere, from each side, it expands as a cylinder and that the flat part does not expand, but it stays constant as the wavefront moves downwards; this idealisation is depicted in Fig.~\ref{wave-drawing}.

If we follow this scheme the area of the expanding wavefront could be calculated as the sum of the area of the contacting surface ($A_0$) four quarter cylinders (one per edge) and four eights of a sphere (one per corner).  If $A_w$ is the area of the wavefront then:
\begin{eqnarray}
A_w&=&A_0+4\frac{\pi R l}{4}+4\frac{4\pi R^2}{8}\\
A_w&=& A_0+\pi R \sqrt{A_0} +2\pi R^2
\end{eqnarray}

As it can be seen, $A_w$=$A_0$ for $R=0$ and it tends to $2\pi R^2$ (the area of an hemisphere) for large values of $R$.  Also, if $A_w$ is independent of $R$ as in our simulations, $A_w$=$A_0$.  With this, we can write the following equation:
\begin{equation}
E_{kf}^*=\frac{E_{k0}T^{R/\lambda}} {A_0+\pi R \sqrt{A_0} +2\pi R^2}.
\end{equation}

Now we need to calculate the velocity that the particles will acquire as the peak of the kinetic energy passes through them:
\begin{eqnarray}
\frac{1}{2}m_p v^2_p\frac{\phi}{\pi r_p^2}&=&\frac{E_{k0}T^{R/\lambda}} {A_0+\pi R \sqrt{A_0} +2\pi R^2}\\
v_p&=&\sqrt{\frac{3}{2}E_{k0}\frac{T^{R/\lambda}}{r_p\rho_p\phi(A_0+\pi R\sqrt{A_0}+2\pi R^2)}},
\label{vparticle}
\end{eqnarray}
where $r_p$ is the particle radius, $\rho_p$ is its density and $\phi$ is the filling fraction of the granular medium.  Notice that in what follows we are assuming that the entire energy of the wave is carried as a Dirac Delta by the peak.  This will provide an upper bound in the effects.
\begin{figure}[h]
\begin{center}
\includegraphics[scale=1]{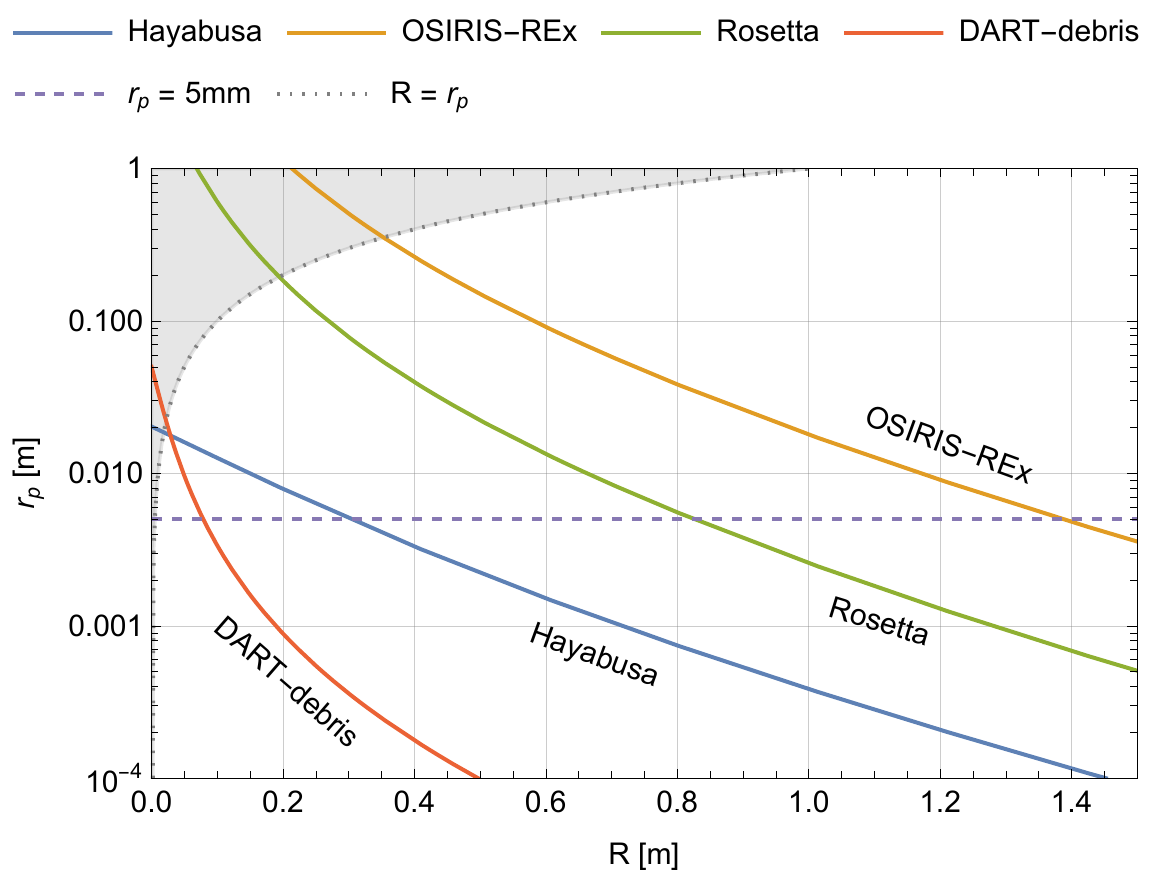}
\caption{The different lines represent the distance at which particles of a certain size could have acquired enough energy to escape the gravitational field of the target bodies upon the impacts of the S/C for each space mission.  The horizontal line represents $r_p=5$ mm and its intersection with the mission lines points to the distance $R$ at which these particles would have enough energy to reach escape velocity.}
\label{escaping-particles}
\end{center}
\end{figure}

{\it The Hayabusa mission}: the spacecraft (S/C) had a mass of 510 kg,  and the dimensions of the core were 1.0m x 1.1m x 1.6m (Hexahedron) \citep{hayabusa-web} and the touchdown velocity was calculated to be $\approx$6.9 cm/s \citep{science_yano}.  The macroporosity of asteroid Itokawa (1-$\phi$) has been calculated at $\approx 41\%$ \citep{fujiwara}.  If we assume a perfect contact between two flat surfaces (as the sampler horn was not a solid structure), we would have a momentum per unit area of $\approx$32 kg m/m$^2$.  Using the equation for the trend lines in Fig.~\ref{scaling} we obtain a $\approx247$ or $211.3$ m/s for the velocity of the peak depending on whether we use a energy ($U_{pk}$) of momentum ($U_{pm}$)scaling respectively.  Note that this velocity could be dependent on the particle size (something we have not tested), and might increase for smaller particles.  The escape velocity for asteroid Itokawa is $\approx$0.2 m/s \citep{engelmann-epsc2017} which means that 1cm particles, density of 3400 kg/m$^3$ \citep{tsuchiyama2011}, up to $\approx$30 cm away from the border of the S/C, could have a high enough velocity to escape.  The larger the particles, the closer they have to be to the S/C to acquire enough energy to escape the gravitational field.  Smaller particles could be farther, but cohesive forces begin to be important for them and some of the energy should be used to break the cohesive bond \citep{scheeres2010}.

{\it The OSIRIS-REx mission}: the S/C of this mission had a mass $\approx$1300 kg \citep{ballouz2021} when the sampling manoeuvre took place and the nominal approaching velocity was $\approx$0.1 m/s.  During the manoeuvre, the TAGSAM (Touch-and-go-Sample-Acquisition-Mechanism) arm did not retract and so the entire mass of the S/C interacted with the surface of the asteroid Bennu for a short while before the gas (to drive the sample into the sample chamber) was fired.  Additionally, the TAGSAM head had an outer ring that made first contact with the asteroid surface and had an area of $\approx$0.046 m$^2$.  Particle density is $\approx$2000 kg/m$^3$ and the escape velocity has a maximum and minimum of 0.178 m/s and 0.00534 m/s respectively \citep{jiang2020}.  Around Osprey, the sampling site, the escape velocity is $\approx$0.09 m/s.  Macroporosity of Bennu has been calculated at 12\% \citep{tricarico2021}.  This means that 1cm up to $\approx$1.39 m from the impact site would have had enough energy to escape Bennu's gravitational field.  The images taken by the spacecraft \citep{walsh2022-tagsam} show that an area of 0.51 $m^2$ around the impact point were disturbed due to the TAGSAM manoeuvre; which implies a radius of $\approx$ 40 cm.  Our result is about 3 times larger; however, given that we are assuming that the entire energy left in the system is carried by the peak, this is not a surprise.   Furthermore, it shows that our results are an upper bound of reality as expected.

{\it The ROSETTA mission}:  The lander of the ROSETTA mission, Philae, had a mass of 97.63 kg and used 3 landing pads \citep{bibring2007,biele2015}.  The landing on asteroid Churiomov-Gerasimenko 67/P had an initial energy of 7.02 $\pm$ 0.37 J before contact and 2.83 $\pm$ 0.22 J afterwards.  Additionally, Philae made contact with the surface with only the side of one landing pad, which leads to a contact area of 225 $\pm$ 40 cm$^2$ \citep{heinisch2019}.  Given that Philae could rotate, it is difficult to use the momentum equation (fig.~\ref{scaling}), but the energy scaling equation results in $U_p\approx 393$ m/s.  The density of the comets is calculated at 532 $\pm$ 7 kg/m$^3$ and its porosity from 70\%-75\% \citep{jorda2016}.  Which means the grain density is $\approx$1773 kg/m$^3$.  The escape velocity of this comet is 0.5 m/s \citep{fulle2010,thomas2015,fulle2016}, which means that 1cm particles up to $\approx$0.83 m from the impact point had enough energy to escape.

In all these cases, the interaction between the S/C and the surface of a small body created a small, but visible crater.  The distances up to which 1cm particles could have acquired enough energy to escape are provided as a rough approximation of the crater size, though it is a very crude way to calculate it.  For the next case, the DART mission, the crater formation process is different and so we will not attempt to approximate its size.  However, some of the particles that form the ejecta field will land on Didymos at relatively slow impact velocities.

{\it The DART mission}: The impact of the S/C will be hypersonic and it will generate an ejecta field.  Part of this ejecta will escape, part will land back on Dimorphos and part will fall on Didymos.  The debris falling on Didymos, up to 1 m/s \citep{yu2018}.  If we suppose that particles as large as 10 cm will impact Didymos, it is possible to do some calculations.  The bulk density of Dimorphos has been calculated to be $\approx$2200.0 kg/m$^3$ \citep{agrusa2022} and so each impact will deliver Ek=73.8 J/m$^2$ on the assumed regolith-covered surface on Didymos.  The porosity of asteroid Didymos is uncertain at the moment of this writing and it has been estimated to be  $\approx$40\% \citep{naidu2020}.  If the escape velocity of Didymos is 1 m/s \citep{michel2016}, each impact will provide enough energy for 1cm particles up to $\approx$8 cm from the borders of the impactors to escape Didymos.

In every case, larger particles could also escape, but they would have to be closer to the impact point.  Particles that do not have enough energy to escape, would still be able to move and delineate a crater.  Fig.~\ref{escaping-particles} shows these results for all the cases discussed above.  We have used equation \ref{vparticle}, equating $v_p$ to the escape velocity of each body to obtain relationships between particle radii and distance from the impact site.   Particles to the right of the mission lines would be too far and will not have enough energy; particles above these lines would be too large.  Particles to the left/above the gray, dotted line ($r_p=R$) cannot exist as they should be closer to the impact site than their radii would allow.  The purple, dashed, horizontal line represents $r_p$=5 mm (1 cm in size/diameter), so the intersection between this line and the mission lines mark the distances at which 1 cm particles would be able to acquire enough kinetic energy to escape their respective gravitational field.

\subsection{High Velocity Impacts}

\paragraph{The DART mission:} The S/C of the DART mission will impact asteroid Dimorphos in a few months as of the writing of this paper.  The impact will take place at $\approx$ 6.58 km/s on the 26$^{th}$ of September, 2022.  The mass of the S/C will be of $\approx$610 kg at impact \citep{cheng2020}; this implies that the kinetic energy delivered will be $\approx 1.32\times 10^{10}$J (not taking into account any remaining fuel) and the radius of the crater generated is predicted be between 5.96-45.1 m depending on the material properties of the asteroid.  The crater depth-to-diameter ratio for this size crater is 0.11 \citep{stopar2017} and so, the depth of the crater in Dimorphos should be between 1.32-9.9 m.  The crater size depends on the material properties of Dimorphos and those have not been established yet; however, we can assume, as \cite{naidu2020}, that Dimorphos and Didymos have the same bulk density and porosity (2200 kg/m$^3$, 40\% respectively).  Given the size of Dimorphos ($\approx$160m in diameter), the distance that the seismic wave would have to travel to reach the antipode to the impact would change minimally.  Additionally, the escape velocity of Dimorphos is $\approx$0.03 m/s \citep{bernauer2020}.

With all this in mind, in the best case scenario when the entirety of the energy of the impact is converted into a seismic wave, 1 cm particles would acquire enough energy to escape to asteroid gravitational field up to 12.5 m and 11 m from the wall of the crater for the small and big craters respectively for the two possible crater sizes.  Particles 17 m away from the crater would have to be in the micro-metre range to have a speed of 0.003 m/s.  This would make it impossible for regolith on the antipode to be lofted.  However, surface particles a few tens of metres outside the crater will be displaced and possibly escape along with the particles forming the ejecta field.

\paragraph{The Haybusa2 mission:}
Impacts such as that executed on asteroid Ryugu by the Hayabusa2 mission are at least supersonic; it should not have caused much of a disturbance on the asteroid as a whole, but just in a small region around the impact site.  The Small Carry-on Impactor (SCI) of the Hayabusa 2 mission had an impact velocity of 2 km/s and a mass of 2 kg.  The escape velocity of asteroid Ryugu is $\approx$0.14 m/s \citep{kikuchi2020} and the density of its regolith is $\approx$ 1282 kg/m$^3$ \citep{yada2022}.  Additionally, the macroporosity of asteroid Ryugu has been calculated to be 14\% and the particle density, 1380 kg/m$^3$ \citep{herbst2021}.  If as in the case for the DART impact, we assume that all the energy is converted into a seismic wave that starts at the edge of the crater, 1 cm particles $\approx$5 m from the edge of the crater would have had enough energy to escape de asteroid and at the 10 m mark, the particle size would have had to be reduced to 1 $\mu$m to reach this speed, but, as said above, at this particle size cohesive forces need to be taken into account.  Please note that we have not taken into account the energy spent in compressing the very porous surface, how porosity affects energy dissipation, or the formation of an ejecta field.  However, \cite{nishiyama2021} has carried out an extended study of the SCI impact and has also concluded that the seismic waves that were generated dissipate rapidly, or energy conversion from impact to seismic waves is small.

\subsection{Other implications}
Another implication of our findings is that seismic shaking over long distances is unlikely to take place on granular asteroids.  If seismic shaking due to small meteorite impacts on small bodies is not produced, or at least not to the extent that has previously been portrayed, the Brazil Nut Effect (BNE) \citep{LPSC_BNE} might not be a possible explanation for {\it global particle size segregation} of the kind observed on asteroid Itokawa \citep{itokawa_miyamoto} by the Haybusa mission.  This conclusion is also supported by what was found by previous studies; \cite{sanchez-lpsc2010} argued that a 2D, self-gravitating aggregate, if continuously impacted in random directions, would show an outer ring in which its particle are size segregated.  This result was also enhanced by the fact that the smaller particles would escape after an energetic collision.  \cite{perera2016} showed that if the particles of a spherical aggregate were provided with a small random velocity (seismic shaking), the aggregate would segregate the particles on its surface by size, but its interior would remain mixed.  These authors made sure that the provided random velocity was not enough for the particles to escape.  \cite{shinbrot2017} have completely discarded the BNE in favour what they have called {\it ballistic sorting}.  At the same time \cite{wright2020} proposed that a single pulse can cause boulders to wind up on top via ballistic sorting if material is lofted by the pulse and so shaking would not be required.  \cite{sautel2021} found that the particles of a simulated asteroid would segregate almost completely by size when short quakes (inversion of the gravitational field) were applied to the system; however, they recognise that future work should focus on a more realistic perturbation method.  Regardless of the true mechanism that is driving the particle segregation that is observed, one thing seems certain, if impacts are the ultimate cause, segregation takes place only on the surface, not the interior.  If this is so, as argued by \cite{sanchez2018}, an asteroid with a weak core could be formed by the size segregation of the particles on its surface.  This would provide a very weak outer shell formed by large particles, a strong middle shell formed by small regolith and dust, and a well mixed interior which would be weaker than the previous shell.  This would form a three-shell structure for a quasi-spherical asteroid.  If the spin rate of this body is subsequently increased by the YORP effect, the core would fail before the other two shells, and at the appropriate size (radius of the core is $\approx$0.7 of the radius of the body), the body would deform towards the shape of a peanut, similar to Itokawa, by breaking the outermost shell, as shown by \cite{sanchez2018}.  This would expose the finer material of the intermediate shell (also mentioned by \cite{walsh2022-tagsam} about asteroid Bennu) in the neck and back regions, leaving the coarser material at both ends of the deformed body.

Though it is difficult to definitively ascertain the strength and structure of asteroid interiors, there are some observations made by the  OSIRIS-REx and Hayabusa2 missions that indicate this configuration.  The analysis of the interior of the craters on asteroid Bennu show that crater morphology changes at $\approx20$ m in diameter.  The floor of craters below this size are generally rock-free and appear to contain an abundance of finer material \citep{bierhaus-lpsc2019,bierhaus-lpsc2020}.  On the other hand, \cite{arakawa2020} mentions that the crater formed by the SCI impact has a central pit, similar to what is observed on crater on the Moon and that this happens because of its layered surface (hard subsurface below cohesionless regolith).  They explain that the surface of Ryugu could have a similar layered structure.  Additionally, assuming that the crater wall represents the subsurface structure, they conclude that the subsurface layer is dominated by regolith with rock sizes smaller than 20 cm.

Putting all this together, the evidence from simulations about impacts on small self-gravitating aggregates predict a layered surface due to the segregation of their particles and escape of the smaller ones.  In this paper, we have presented evidence that suggests that granular matter at extremely low pressure can dissipate energy very rapidly.  Which would imply that if a layered surface exists in real asteroids, this layer would be up to a couple of tens of meters thick.  The evidence from recent space missions show that the particles that form the surface of small, granular asteroids are segregated as observed in the interior of natural and artificially formed craters.  This finding, in our view, seems to confirm the predictions made by the authors cited above.

We should also point out that if the efficiency of seismic wave transmission is as low as we have been able to measure with our simulations, the fluidisation of the entirety of the body of a granular asteroid due to micro-meteorite impacts \citep{tanga2009} seems unlikely.  This is, major shape deformation due to small micro-meteorite impacts would not be expected.  Surely, if the impact is large enough to cause the complete disruption of the asteroid a new shape would emerge, but this would then be a new asteroid formed by the remnants of the parent bodies.

Finally, if, as we have shown here, the speed of sound in granular materials at extremely low pressure depends on the velocity of the impacts, care should be taken when classifying whether an impact is truly at hyper- or supersonic velocities.  This will possibly have repercussions on the numerical and theoretical tools that are used to study them.

\section{Conclusions}

In this paper we have studied how seismic waves generated by a single pulse, used as a proxy for a low velocity impact, are transmitted in granular media with a view to apply our findings to the asteroid environment where the confining pressure goes from almost zero to just a few tens of Pascals.

The study has been carried out through simulation, using a Soft-Sphere DEM code using perfect spheres as a proxy to real grains.  We have tested confining pressures from 0.1- 50000 Pa and impact/compression speeds from 0.001-5 m/s.

We have shown that the speed of seismic waves at low enough confining pressures depends not only on the confining pressure ($P_c$), but also on the pressure that is produced by the wave itself ($P_i$).  Also, that wave speed ($v_w$) of compression waves depends approximately on $P_s^{1/6}$, where $P_s=P_c+P_i$ is the stable pressure on the section of the material through which the wavefront has already passed.

Our simulations show that the velocity of transmission of the peak ($U_p$) pressure depends not only on the confining pressure, but also on the impact that produced it.  Also, within the regime that was investigated, $U_p$ can be scaled either through the momentum or the kinetic energy of the impact.  More importantly, and in view of the immediate application of this research, our simulations have shown that pulse pressure attenuates very rapidly and its energy is quickly dissipated.  With these results in mind, we have linked our results to the possible seismic behaviour of granular asteroids (mind the caveats) and also tried to provide some insight into the possible surface dynamics that could have taken place on different small solar system bodies with which a few spacecrafts have interacted or will interact.

In addition to this, and given how greatly energy can be dissipated in a granular medium even at extremely low confining pressures, we have found that impacts on the surface of small granular asteroids would have affected only their surface (possibly down to a few tens of metres in depth).  This implies that all the segregation phenomena, of which the morphology of observed craters (on asteroids Bennu and Ryugu) provide evidence, would be restricted to this external shell.  This would in turn imply a layered structure is formed, at the very least, by an outer layer formed by large grains and mostly cohesionless, a middle layer formed by smaller grains that would have greater cohesive strength and, an inner core, undisturbed by the impacts (not segregated), which would be structurally weaker than the middle layer.  This structure, as explained above and argued in previous research \citep{sanchez2018} could explain the segregation observed in asteroid Itokawa.

\begin{acknowledgments}

The authors acknowledge financial support from NASA Grant 80NSSC18K0491.  PS would like to thank Beau Bierhaus (Lockheed Martin Aerospace) for his help on the discussion of the findings of the OSIRIS-REx mission on Bennu, and Hongyang Chenk (University of Twente) for the helpful discussions and for commenting on earlier versions of this manuscript.
\end{acknowledgments}

\bibliography{/Users/paul/Documents/UCB/Meeting/psbib}

\begin{thebibliography}{}
\expandafter\ifx\csname natexlab\endcsname\relax\def\natexlab#1{#1}\fi

\bibitem[{{Agrusa, H. F.} {et~al.}(2022){Agrusa, H. F.}, {Ballouz, R.}, {Meyer,
  A. J.}, {Tasev, E.}, {Noiset, G.}, {Karatekin, \"O.}, {Michel, P.},
  {Richardson, D. C.}, \& {Hirabayashi, M.}}]{agrusa2022}
{Agrusa, H. F.}, {Ballouz, R.}, {Meyer, A. J.}, {et~al.} 2022, A\&A, 664, L3

\bibitem[{Anthony {et~al.}(2004)Anthony, Hoyle, \& Ding}]{janthony2004}
Anthony, S.~J., Hoyle, W., \& Ding, Y. 2004, Granular Materials: Fundamentals
  and Applications (Cambridge, UK: The Royal Society of Chemistry), 5--7

\bibitem[{Arakawa {et~al.}(2017)Arakawa, Wada, Saiki, Kadono, Takagi, Shirai,
  Okamoto, Yano, Hayakawa, Nakazawa, Hirata, Kobayashi, Michel, Jutzi, Imamura,
  Ogawa, Sakatani, Iijima, Honda, Ishibashi, Hayakawa, \& Sawada}]{arakawa2017}
Arakawa, M., Wada, K., Saiki, T., {et~al.} 2017, Space Science Reviews, 208,
  187

\bibitem[{Arakawa {et~al.}(2020)Arakawa, Saiki, Wada, Ogawa, Kadono, Shirai,
  Sawada, Ishibashi, Honda, Sakatani, Iijima, Okamoto, Yano, Takagi, Hayakawa,
  Michel, Jutzi, Shimaki, Kimura, Mimasu, Toda, Imamura, Nakazawa, Hayakawa,
  Sugita, Morota, Kameda, Tatsumi, Cho, Yoshioka, Yokota, Matsuoka, Yamada,
  Kouyama, Honda, Tsuda, Watanabe, Yoshikawa, Tanaka, Terui, Kikuchi,
  Yamaguchi, Ogawa, Ono, Yoshikawa, Takahashi, Takei, Fujii, Takeuchi,
  Yamamoto, Okada, Hirose, Hosoda, Mori, Shimada, Soldini, Tsukizaki, Iwata,
  Ozaki, Abe, Namiki, Kitazato, Tachibana, Ikeda, Hirata, Hirata, Noguchi, \&
  Miura}]{arakawa2020}
Arakawa, M., Saiki, T., Wada, K., {et~al.} 2020, Science, 368, 67

\bibitem[{{Asphaug} {et~al.}(2001){Asphaug}, {King}, {Swift}, \&
  {Merrifield}}]{LPSC_BNE}
{Asphaug}, E., {King}, P.~J., {Swift}, M.~R., \& {Merrifield}, M.~R. 2001, in
  Lunar and Planetary Inst. Technical Report, Vol.~32, Lunar and Planetary
  Institute Science Conference Abstracts, 1708--+

\bibitem[{Auma\^itre {et~al.}(2018)Auma\^itre, Behringer, Cazaubiel, Cl\'ement,
  Crassous, Durian, Falcon, Fauve, Fischer, Garcimart\'in, Garrabos, Hou, Jia,
  Lecoutre, Luding, Maza, Noirhomme, Opsomer, Palencia, P\"oschel, Schockmel,
  Sperl, Stannarius, Vandewalle, \& Yu}]{aumaitre2018}
Auma\^itre, S., Behringer, R.~P., Cazaubiel, A., {et~al.} 2018, Review of
  Scientific Instruments, 89, 075103

\bibitem[{Ballouz {et~al.}(2021)Ballouz, Walsh, Sánchez, Holsapple, Michel,
  Scheeres, Zhang, Richardson, Barnouin, Nolan, Bierhaus, Connolly, Schwartz,
  Çelik, Baba, \& Lauretta}]{ballouz2021}
Ballouz, R.-L., Walsh, K.~J., Sánchez, P., {et~al.} 2021, Monthly Notices of
  the Royal Astronomical Society, 507, 5087

\bibitem[{Bernauer {et~al.}(2020)Bernauer, Garcia, Murdoch, Dehant, Sollberger,
  Schmelzbach, St{\"a}hler, Wassermann, Igel, Cadu, Mimoun, Ritter, Filice,
  Karatekin, Ferraioli, Robertsson, Giardini, Lecamp, Guattari, Bonnefois, \&
  de~Raucourt}]{bernauer2020}
Bernauer, F., Garcia, R.~F., Murdoch, N., {et~al.} 2020, Earth, Planets and
  Space, 72, 191

\bibitem[{Bibring {et~al.}(2007)Bibring, Rosenbauer, Boehnhardt, Ulamec, Biele,
  Espinasse, Feuerbacher, Gaudon, Hemmerich, Kletzkine, Moura, Mugnuolo,
  Nietner, P{\"a}tz, Roll, Scheuerle, Szeg{\"o}, Wittmann, Office, \&
  Team9}]{bibring2007}
Bibring, J.~P., Rosenbauer, H., Boehnhardt, H., {et~al.} 2007, Space Science
  Reviews, 128, 205

\bibitem[{Biele {et~al.}(2015)Biele, Ulamec, Maibaum, Roll, Witte, Jurado,
  Mu{\~n}oz, Arnold, Auster, Casas, Faber, Fantinati, Finke, Fischer, Geurts,
  G{\"u}ttler, Heinisch, Herique, Hviid, Kargl, Knapmeyer, Knollenberg, Kofman,
  K{\"o}mle, K{\"u}hrt, Lommatsch, Mottola, Pardo~de Santayana, Remetean,
  Scholten, Seidensticker, Sierks, \& Spohn}]{biele2015}
Biele, J., Ulamec, S., Maibaum, M., {et~al.} 2015, Science, 349,
  http://science.sciencemag.org/content/349/6247/aaa9816.full.pdf

\bibitem[{Bierhaus {et~al.}(2019)Bierhaus, Barnouin, Walsh, Daly, Pajola,
  Jawin, McCoy, Connolly, Dellagiustina, Rizk, {et~al.}}]{bierhaus-lpsc2019}
Bierhaus, E., Barnouin, O., Walsh, K., {et~al.} 2019, EPSC, 2019, EPSC

\bibitem[{Bierhaus {et~al.}(2020)Bierhaus, Trang, Barnouin, Daly, Walsh, Daly,
  DellaGiustina, Michel, Susorney, Johnson, {et~al.}}]{bierhaus-lpsc2020}
Bierhaus, E., Trang, D., Barnouin, O., {et~al.} 2020, LPI, 2156

\bibitem[{Biswas {et~al.}(2003)Biswas, S\'anchez, Swift, \& King}]{bis}
Biswas, P., S\'anchez, P., Swift, M., \& King, P. 2003, Phys. Rev. E, 68,
  050301(R)

\bibitem[{Bottke {et~al.}(1994)Bottke, Nolan, Greenberg, \&
  Kolvoord}]{bottke1994}
Bottke, W.~F., Nolan, M.~C., Greenberg, R., \& Kolvoord, R.~A. 1994, Icarus,
  107, 255

\bibitem[{Cheng {et~al.}(2020)Cheng, Stickle, Fahnestock, Dotto, {Della Corte},
  Chabot, \& Rivkin}]{cheng2020}
Cheng, A.~F., Stickle, A.~M., Fahnestock, E.~G., {et~al.} 2020, Icarus, 352,
  113989

\bibitem[{Cundall(1971)}]{cundall1971}
Cundall, P. 1971, in Proceedings of the International Symposium on Rock
  Mechanics, Vol.~1 (Nancy: -), 129--136

\bibitem[{Cundall \& Hart(1992)}]{cundall1992}
Cundall, P.~A., \& Hart, R.~D. 1992, Engineering Computations, 9, 101

\bibitem[{DellaGiustina {et~al.}(2019)DellaGiustina, Emery, Golish, Rozitis,
  Bennett, Burke, Ballouz, Becker, Christensen, Drouet~d'Aubigny, Hamilton,
  Reuter, Rizk, Simon, Asphaug, Bandfield, Barnouin, Barucci, Bierhaus, Binzel,
  Bottke, Bowles, Campins, Clark, Clark, Connolly, Daly, Leon, Delbo',
  Deshapriya, Elder, Fornasier, Hergenrother, Howell, Jawin, Kaplan, Kareta,
  Le~Corre, Li, Licandro, Lim, Michel, Molaro, Nolan, Pajola, Popescu, Garcia,
  Ryan, Schwartz, Shultz, Siegler, Smith, Tatsumi, Thomas, Walsh, Wolner, Zou,
  Lauretta, Highsmith, Small, Vokrouhlick{\'y}, Bowles, Brown, Donaldson~Hanna,
  Warren, Brunet, Chicoine, Desjardins, Gaudreau, Haltigin, Millington-Veloza,
  Rubi, Aponte, Gorius, Lunsford, Allen, Grindlay, Guevel, Hoak, Hong,
  Schrader, Bayron, Golubov, S{\'a}nchez, Stromberg, Hirabayashi, Hartzell,
  Oliver, Rascon, Harch, Joseph, Squyres, Richardson, Emery, McGraw, Ghent,
  Binzel, Asad, Johnson, Philpott, Susorney, Cloutis, Hanna, Connolly, Ciceri,
  Hildebrand, Ibrahim, Breitenfeld, Glotch, Rogers, Clark, Ferrone, Thomas,
  Campins, Fernandez, Chang, Cheuvront, Trang, Tachibana, Yurimoto, Brucato,
  Poggiali, Pajola, Dotto, Epifani, Crombie, Lantz, Izawa, de~Leon, Licandro,
  Garcia, Clemett, Thomas-Keprta, Van~wal, Yoshikawa, Bellerose, Bhaskaran,
  Boyles, Chesley, Elder, Farnocchia, Harbison, Kennedy, Knight,
  Martinez-Vlasoff, Mastrodemos, McElrath, Owen, Park, Rush, Swanson,
  Takahashi, Velez, Yetter, Thayer, Adam, Antreasian, Bauman, Bryan, Carcich,
  Corvin, Geeraert, Hoffman, Leonard, Lessac-Chenen, Levine, McAdams, McCarthy,
  Nelson, Page, Pelgrift, Sahr, Stakkestad, Stanbridge, Wibben, Williams,
  Williams, Wolff, Hayne, Kubitschek, Barucci, Deshapriya, Fornasier,
  Fulchignoni, Hasselmann, Merlin, Praet, Bierhaus, Billett, Boggs, Buck,
  Carlson-Kelly, Cerna, Chaffin, Church, Coltrin, Daly, Deguzman, Dubisher,
  Eckart, Ellis, Falkenstern, Fisher, Fisher, Fleming, Fortney, Francis,
  Freund, Gonzales, Haas, Hasten, Hauf, Hilbert, Howell, Jaen, Jayakody,
  Jenkins, Johnson, Lefevre, Ma, Mario, Martin, May, McGee, Miller, Miller,
  Miller, Mirfakhrai, Muhle, Norman, Olds, Parish, Ryle, Schmitzer, Sherman,
  Skeen, Susak, Sutter, Tran, Welch, Witherspoon, Wood, Zareski,
  Arvizu-Jakubicki, Asphaug, Audi, Ballouz, Bandrowski, Becker, Becker,
  Bendall, Bennett, Bloomenthal, Blum, Boynton, Brodbeck, Burke, Chojnacki,
  Colpo, Contreras, Cutts, Drouet~d'Aubigny, Dean, DellaGiustina, Diallo,
  Drinnon, Drozd, Enos, Enos, Fellows, Ferro, Fisher, Fitzgibbon, Fitzgibbon,
  Forelli, Forrester, Galinsky, Garcia, Gardner, Golish, Habib, Hamara,
  Hammond, Hanley, Harshman, Hergenrother, Herzog, Hill, Hoekenga, Hooven,
  Howell, Huettner, Janakus, Jones, Kareta, Kidd, Kingsbury, Balram-Knutson,
  Koelbel, \& Team}]{dellagiustina2019}
DellaGiustina, D.~N., Emery, J.~P., Golish, D.~R., {et~al.} 2019, Nature
  Astronomy, 3, 341

\bibitem[{Digby(1981)}]{digby1981}
Digby, P.~J. 1981, Journal of Applied Mechanics, 48, 803

\bibitem[{Engelmann {et~al.}(2017)Engelmann, W{\"u}nnemann, Luther, \&
  Zhu}]{engelmann-epsc2017}
Engelmann, J., W{\"u}nnemann, K., Luther, R., \& Zhu, M.-H. 2017, in European
  Planetary Science Congress, EPSC2017--251

\bibitem[{Fonseka {et~al.}(2022)Fonseka, Awasthi, Lambros, \&
  Geubelle}]{fonseka2022}
Fonseka, R. D. J.~I., Awasthi, A., Lambros, J., \& Geubelle, P.~H. 2022,
  Journal of Applied Mechanics, 89,
  https://asmedigitalcollection.asme.org/appliedmechanics/article-pdf/89/5/051003/6839128/jam\_89\_5\_051003.pdf,
  051003

\bibitem[{{Fujiwara} {et~al.}(2006{\natexlab{a}}){Fujiwara}, {Kawaguchi},
  {Yeomans}, {Abe}, {Mukai}, {Okada}, {Saito}, {Yano}, {Yoshikawa}, {Scheeres},
  {Barnouin-Jha}, {Cheng}, {Demura}, {Gaskell}, {Hirata}, {Ikeda}, {Kominato},
  {Miyamoto}, {Nakamura}, {Nakamura}, {Sasaki}, \& {Uesugi}}]{science_fujiwara}
{Fujiwara}, A., {Kawaguchi}, J., {Yeomans}, D.~K., {et~al.} 2006{\natexlab{a}},
  Science, 312, 1330

\bibitem[{{Fujiwara} {et~al.}(2006{\natexlab{b}}){Fujiwara}, {Kawaguchi},
  {Yeomans}, {Abe}, {Mukai}, {Okada}, {Saito}, {Yano}, {Yoshikawa}, {Scheeres},
  {Barnouin-Jha}, {Cheng}, {Demura}, {Gaskell}, {Hirata}, {Ikeda}, {Kominato},
  {Miyamoto}, {Nakamura}, {Nakamura}, {Sasaki}, \& {Uesugi}}]{fujiwara}
---. 2006{\natexlab{b}}, Science, 312, 1330

\bibitem[{Fulle {et~al.}(2016)Fulle, Marzari, Corte, Fornasier, Sierks,
  Rotundi, Barbieri, Lamy, Rodrigo, Koschny, Rickman, Keller,
  L{\'{o}}pez-Moreno, Accolla, Agarwal, A'Hearn, Altobelli, Barucci, Bertaux,
  Bertini, Bodewits, Bussoletti, Colangeli, Cosi, Cremonese, Crifo, Deppo,
  Davidsson, Debei, Cecco, Esposito, Ferrari, Giovane, Gustafson, Green,
  Groussin, Grün, Gutierrez, Güttler, Herranz, Hviid, Ip, Ivanovski,
  Jer{\'{o}}nimo, Jorda, Knollenberg, Kramm, Kührt, Küppers, Lara, Lazzarin,
  Leese, L{\'{o}}pez-Jim{\'{e}}nez, Lucarelli, Epifani, McDonnell, Mennella,
  Molina, Morales, Moreno, Mottola, Naletto, Oklay, Ortiz, Palomba, Palumbo,
  Perrin, Rietmeijer, Rodr{\'{\i}}guez, Sordini, Thomas, Tubiana, Vincent,
  Weissman, Wenzel, Zakharov, \& Zarnecki}]{fulle2016}
Fulle, M., Marzari, F., Corte, V.~D., {et~al.} 2016, The Astrophysical Journal,
  821, 19

\bibitem[{{Fulle, M.} {et~al.}(2010){Fulle, M.}, {Colangeli, L.}, {Agarwal,
  J.}, {Aronica, A.}, {Della Corte, V.}, {Esposito, F.}, {Gr\"un, E.},
  {Ishiguro, M.}, {Ligustri, R.}, {Lopez Moreno, J. J.}, {Mazzotta Epifani,
  E.}, {Milani, G.}, {Moreno, F.}, {Palumbo, P.}, {Rodr\'{\i}guez G\'omez, J.},
  \& {Rotundi, A.}}]{fulle2010}
{Fulle, M.}, {Colangeli, L.}, {Agarwal, J.}, {et~al.} 2010, A\&A, 522, A63

\bibitem[{Goddard(1990)}]{goddard1990}
Goddard, J.~D. 1990, Proceedings of the Royal Society of London. Series A:
  Mathematical and Physical Sciences, 430, 105

\bibitem[{G\'omez {et~al.}(2012)G\'omez, Turner, van Hecke, \&
  Vitelli}]{gomez2012}
G\'omez, L.~R., Turner, A.~M., van Hecke, M., \& Vitelli, V. 2012, Phys. Rev.
  Lett., 108, 058001

\bibitem[{Hara(1935)}]{hara1935}
Hara, G. 1935, Elektr. Nachr. Techn, 12, 191

\bibitem[{{Heinisch, P.} {et~al.}(2019){Heinisch, P.}, {Auster, H.-U.},
  {Gundlach, B.}, {Blum, J.}, {G\"uttler, C.}, {Tubiana, C.}, {Sierks, H.},
  {Hilchenbach, M.}, {Biele, J.}, {Richter, I.}, \& {Glassmeier, K.
  H.}}]{heinisch2019}
{Heinisch, P.}, {Auster, H.-U.}, {Gundlach, B.}, {et~al.} 2019, A\&A, 630, A2

\bibitem[{Herbst {et~al.}(2021)Herbst, Greenwood, \& Yap}]{herbst2021}
Herbst, W., Greenwood, J.~P., \& Yap, T.~E. 2021, The Planetary Science
  Journal, 2, 110

\bibitem[{Herrmann \& Luding(1998)}]{herr1}
Herrmann, H., \& Luding, S. 1998, Continuum Mechanics and Thermodynamics, 10,
  189, 10.1007/s001610050089

\bibitem[{Hostler \& Brennen(2005)}]{hostler2005}
Hostler, S.~R., \& Brennen, C.~E. 2005, Physical Review E, 72, 031303

\bibitem[{ISAS(2022)}]{hayabusa-web}
ISAS. 2022, Hayabusa website,
  \url{https://www.isas.jaxa.jp/en/missions/spacecraft/past/hayabusa.html}, , ,
  accessed: 2022-03-11

\bibitem[{Jiang \& Schmidt(2020)}]{jiang2020}
Jiang, Y., \& Schmidt, J. 2020, Heliyon, 6, e05275

\bibitem[{Jorda {et~al.}(2016)Jorda, Gaskell, Capanna, Hviid, Lamy, Ďurech,
  Faury, Groussin, Gutiérrez, Jackman, Keihm, Keller, Knollenberg, Kührt,
  Marchi, Mottola, Palmer, Schloerb, Sierks, Vincent, A’Hearn, Barbieri,
  Rodrigo, Koschny, Rickman, Barucci, Bertaux, Bertini, Cremonese, {Da Deppo},
  Davidsson, Debei, {De Cecco}, Fornasier, Fulle, Güttler, Ip, Kramm,
  K\"uppers, Lara, Lazzarin, {Lopez Moreno}, Marzari, Naletto, Oklay, Thomas,
  Tubiana, \& Wenzel}]{jorda2016}
Jorda, L., Gaskell, R., Capanna, C., {et~al.} 2016, Icarus, 277, 257

\bibitem[{Kikuchi {et~al.}(2020)Kikuchi, Terui, Ogawa, Saiki, Ono, Yoshikawa,
  Takei, Mimasu, Ikeda, Sawada, Wal, Sugita, Watanabe, \& Tsuda}]{kikuchi2020}
Kikuchi, S., Terui, F., Ogawa, N., {et~al.} 2020, Journal of Spacecraft and
  Rockets, 57, 1033

\bibitem[{Langlois \& Jia(2015)}]{langlois2015}
Langlois, V., \& Jia, X. 2015, Phys. Rev. E, 91, 022205

\bibitem[{Lauretta {et~al.}(2012)Lauretta, Barucci, Bierhaus, Brucato, Campins,
  Christensen, Clark, Connolly, Dotto, Dworkin, {et~al.}}]{lauretta2012osiris}
Lauretta, D., Barucci, M., Bierhaus, E., {et~al.} 2012, in Comets and Meteors
  2012 Conference No. GSFC. ABS. 6358.2012

\bibitem[{Luding(2008)}]{luding2008}
Luding, S. 2008, Particuology, 6, 501, simulation and Modeling of Particulate
  Systems

\bibitem[{Makse {et~al.}(1999)Makse, Gland, Johnson, \& Schwartz}]{makse1999}
Makse, H.~A., Gland, N., Johnson, D.~L., \& Schwartz, L.~M. 1999, Phys. Rev.
  Lett., 83, 5070

\bibitem[{Michel {et~al.}(2016)Michel, Cheng, Küppers, Pravec, Blum, Delbo,
  Green, Rosenblatt, Tsiganis, Vincent, Biele, Ciarletti, Hérique, Ulamec,
  Carnelli, Galvez, Benner, Naidu, Barnouin, Richardson, Rivkin, Scheirich,
  Moskovitz, Thirouin, Schwartz, {Campo Bagatin}, \& Yu}]{michel2016}
Michel, P., Cheng, A., Küppers, M., {et~al.} 2016, Advances in Space Research,
  57, 2529

\bibitem[{Michikami {et~al.}(2008)Michikami, Nakamura, Hirata, Gaskell,
  Nakamura, Honda, Honda, Hiraoka, Saito, Demura, {et~al.}}]{itokawa_boulders}
Michikami, T., Nakamura, A., Hirata, N., {et~al.} 2008, Earth, Planets, and
  Space, 60, 13

\bibitem[{{Miyamoto} {et~al.}(2007){Miyamoto}, {Yano}, {Scheeres}, {Abe},
  {Barnouin-Jha}, {Cheng}, {Demura}, {Gaskell}, {Hirata}, {Ishiguro},
  {Michikami}, {Nakamura}, {Nakamura}, {Saito}, \& {Sasaki}}]{itokawa_miyamoto}
{Miyamoto}, H., {Yano}, H., {Scheeres}, D.~J., {et~al.} 2007, Science, 316,
  1011

\bibitem[{Naidu {et~al.}(2020)Naidu, Benner, Brozovic, Nolan, Ostro, Margot,
  Giorgini, Hirabayashi, Scheeres, Pravec, Scheirich, Magri, \&
  Jao}]{naidu2020}
Naidu, S., Benner, L., Brozovic, M., {et~al.} 2020, Icarus, 348, 113777

\bibitem[{Nishiyama {et~al.}(2021)Nishiyama, Kawamura, Namiki, Fernando, Leng,
  Onodera, Sugita, Saiki, Imamura, Takagi, Yano, Hayakawa, Okamoto, Sawada,
  Tsuda, Ogawa, Nakazawa, \& Iijima}]{nishiyama2021}
Nishiyama, G., Kawamura, T., Namiki, N., {et~al.} 2021, Journal of Geophysical
  Research: Planets, in press

\bibitem[{Owens \& Daniels(2011)}]{owens2011}
Owens, E.~T., \& Daniels, K.~E. 2011, {EPL} (Europhysics Letters), 94, 54005

\bibitem[{Perera {et~al.}(2016)Perera, Jackson, Asphaug, \&
  Ballouz}]{perera2016}
Perera, V., Jackson, A.~P., Asphaug, E., \& Ballouz, R.-L. 2016, Icarus, 278,
  194

\bibitem[{Quillen {et~al.}(2022)Quillen, Neiderbach, Suo, South, Wright,
  Skerrett, Sánchez, Cúñez, Miklavcic, \& Askari}]{quillen2022}
Quillen, A., Neiderbach, M., Suo, B., {et~al.} 2022, Icarus, 386, 115139

\bibitem[{Richardson {et~al.}(2009)Richardson, Michel, Walsh, \&
  Flynn}]{richardson2009}
Richardson, D., Michel, P., Walsh, K., \& Flynn, K. 2009, Planetary and Space
  Science, 57, 183 , catastrophic Disruption in the Solar SystemVII Workshop on
  Catastrophic Disruption in the Solar System

\bibitem[{Roberts {et~al.}(2021)Roberts, Barnouin, Daly, Walsh, Nolan, Daly,
  Michel, Zhang, Perry, Neumann, Seabrook, Gaskell, Palmer, Weirich, Watanabe,
  Hirata, Hirata, Sugita, Scheeres, McMahon, \& Lauretta}]{roberts2021}
Roberts, J., Barnouin, O., Daly, M., {et~al.} 2021, Planetary and Space
  Science, 204, 105268

\bibitem[{Ru{\'i}z-Botello {et~al.}(2016)Ru{\'i}z-Botello, Castellanos, \&
  Tournat}]{ruiz2016}
Ru{\'i}z-Botello, F., Castellanos, A., \& Tournat, V. 2016, Ultrasonics, 69,
  193

\bibitem[{S{\'a}nchez \& Scheeres(2021{\natexlab{a}})}]{sanchez-lpsc2021}
S{\'a}nchez, D., \& Scheeres, D. 2021{\natexlab{a}}, in Lunar and Planetary
  Science Conference No. 2548, 1850

\bibitem[{S\'anchez \& Scheeres(2012)}]{sanchez2012}
S\'anchez, D.~P., \& Scheeres, D.~J. 2012, Icarus, 218, 876

\bibitem[{{S{\'a}nchez} \& {Scheeres}(2009)}]{sanchez-lpsc2009}
{S{\'a}nchez}, P., \& {Scheeres}, D.~J. 2009, in Lunar and Planetary Inst.
  Technical Report, Vol.~40, Lunar and Planetary Institute Science Conference
  Abstracts, 2228--+

\bibitem[{S\'anchez \& Scheeres(2011)}]{sanchez2011}
S\'anchez, P., \& Scheeres, D.~J. 2011, The Astrophysical Journal, 727, 120

\bibitem[{S\'anchez \& Scheeres(2014)}]{sanchez2014}
---. 2014, Meteoritics \& Planetary Science, 49, 788

\bibitem[{S{\'a}nchez \& Scheeres(2016)}]{sanchez2016}
S{\'a}nchez, P., \& Scheeres, D.~J. 2016, Icarus, 271, 453

\bibitem[{S{\'a}nchez \& Scheeres(2018)}]{sanchez2018}
---. 2018, Planetary and Space Science, 157, 39

\bibitem[{{S\'anchez} \& {Scheeres}(2020)}]{sanchez-lpsc2020}
{S\'anchez}, P., \& {Scheeres}, D.~J. 2020, in Lunar and Planetary Science
  Conference, Lunar and Planetary Science Conference, 1685

\bibitem[{S{\'a}nchez \& Scheeres(2021{\natexlab{b}})}]{sanchez-pg2021}
S{\'a}nchez, P., \& Scheeres, D.~J. 2021{\natexlab{b}}, in EPJ Web of
  Conferences, Vol. 249, EDP Sciences, 13001

\bibitem[{{S\'anchez} {et~al.}(2010){S\'anchez}, {Scheeres}, \&
  {Swift}}]{sanchez-lpsc2010}
{S\'anchez}, P., {Scheeres}, D.~J., \& {Swift}, M.~R. 2010, in Lunar and
  Planetary Inst. Technical Report, Vol.~41, Lunar and Planetary Institute
  Science Conference Abstracts, 2634--+

\bibitem[{S\'{a}nchez {et~al.}(2004)S\'{a}nchez, Swift, \& King}]{sanchez}
S\'{a}nchez, P., Swift, M.~R., \& King, P.~J. 2004, Physical Review Letters,
  93, 184302

\bibitem[{Sautel {et~al.}(2021)Sautel, Lecomte, \& Taberlet}]{sautel2021}
Sautel, J., Lecomte, C.-E., \& Taberlet, N. 2021, Phys. Rev. E, 103, 022901

\bibitem[{Scheeres {et~al.}(2010)Scheeres, Hartzell, S\'anchez, \&
  Swift}]{scheeres2010}
Scheeres, D., Hartzell, C., S\'anchez, P., \& Swift, M. 2010, Icarus, 210, 968

\bibitem[{Schultz(1995)}]{schultz1995}
Schultz, R.~A. 1995, Rock Mechanics and Rock Engineering, 28, 1

\bibitem[{Shinbrot {et~al.}(2017)Shinbrot, Sabuwala, Siu, Vivar~Lazo, \&
  Chakraborty}]{shinbrot2017}
Shinbrot, T., Sabuwala, T., Siu, T., Vivar~Lazo, M., \& Chakraborty, P. 2017,
  Phys. Rev. Lett., 118, 111101

\bibitem[{Silbert {et~al.}(2001)Silbert, Erta\ifmmode~\mbox{\c{s}}\else
  \c{s}\fi{}, Grest, Halsey, Levine, \& Plimpton}]{silbert2001}
Silbert, L.~E., Erta\ifmmode~\mbox{\c{s}}\else \c{s}\fi{}, D., Grest, G.~S.,
  {et~al.} 2001, Phys. Rev. E, 64, 051302

\bibitem[{Stopar {et~al.}(2017)Stopar, Robinson, Barnouin, McEwen, Speyerer,
  Henriksen, \& Sutton}]{stopar2017}
Stopar, J.~D., Robinson, M.~S., Barnouin, O.~S., {et~al.} 2017, Icarus, 298,
  34, lunar Reconnaissance Orbiter - Part III

\bibitem[{Tachibana {et~al.}(2022)Tachibana, Sawada, Okazaki, Takano, Sakamoto,
  Miura, Okamoto, Yano, Yamanouchi, Michel, Zhang, Schwartz, Thuillet,
  Yurimoto, Nakamura, Noguchi, Yabuta, Naraoka, Tsuchiyama, Imae, Kurosawa,
  Nakamura, Ogawa, Sugita, Morota, Honda, Kameda, Tatsumi, Cho, Yoshioka,
  Yokota, Hayakawa, Matsuoka, Sakatani, Yamada, Kouyama, Suzuki, Honda,
  Yoshimitsu, Kubota, Demura, Yada, Nishimura, Yogata, Nakato, Yoshitake,
  Suzuki, Furuya, Hatakeda, Miyazaki, Kumagai, Okada, Abe, Usui, Ireland,
  Fujimoto, Yamada, Arakawa, Connolly, Fujii, Hasegawa, Hirata, Hirata, Hirose,
  Hosoda, Iijima, Ikeda, Ishiguro, Ishihara, Iwata, Kikuchi, Kitazato,
  Lauretta, Libourel, Marty, Matsumoto, Michikami, Mimasu, Miura, Mori,
  Nakamura-Messenger, Namiki, Nguyen, Nittler, Noda, Noguchi, Ogawa, Ono,
  Ozaki, Senshu, Shimada, Shimaki, Shirai, Soldini, Takahashi, Takei, Takeuchi,
  Tsukizaki, Wada, Yamamoto, Yoshikawa, Yumoto, Zolensky, Nakazawa, Terui,
  Tanaka, Saiki, Yoshikawa, Watanabe, \& Tsuda}]{tachibana2022}
Tachibana, S., Sawada, H., Okazaki, R., {et~al.} 2022, Science, 375, 1011

\bibitem[{Tanga {et~al.}(2009)Tanga, Comito, Paolicchi, Hestroffer, Cellino,
  Dell'Oro, Richardson, Walsh, \& Delbo}]{tanga2009}
Tanga, P., Comito, C., Paolicchi, P., {et~al.} 2009, The Astrophysical Journal
  Letters, 706, L197

\bibitem[{Tell {et~al.}(2020)Tell, Drei\ss{}igacker, Tchapnda, Yu, \&
  Sperl}]{tell2020}
Tell, K., Drei\ss{}igacker, C., Tchapnda, A.~C., Yu, P., \& Sperl, M. 2020,
  Review of Scientific Instruments, 91, 033906

\bibitem[{{Thomas} \& {Robinson}(2005)}]{thomas_eros_craters}
{Thomas}, P.~C., \& {Robinson}, M.~S. 2005, Nature, 436, 366

\bibitem[{{Thomas, N.} {et~al.}(2015){Thomas, N.}, {Davidsson, B.}, {El-Maarry,
  M. R.}, {Fornasier, S.}, {Giacomini, L.}, {Gracia-Bern\'a, A. G.}, {Hviid, S.
  F.}, {Ip, W.-H.}, {Jorda, L.}, {Keller, H. U.}, {Knollenberg, J.}, {K\"uhrt,
  E.}, {La Forgia, F.}, {Lai, I. L.}, {Liao, Y.}, {Marschall, R.}, {Massironi,
  M.}, {Mottola, S.}, {Pajola, M.}, {Poch, O.}, {Pommerol, A.}, {Preusker, F.},
  {Scholten, F.}, {Su, C. C.}, {Wu, J. S.}, {Vincent, J.-B.}, {Sierks, H.},
  {Barbieri, C.}, {Lamy, P. L.}, {Rodrigo, R.}, {Koschny, D.}, {Rickman, H.},
  {A\'{}Hearn, M. F.}, {Barucci, M. A.}, {Bertaux, J.-L.}, {Bertini, I.},
  {Cremonese, G.}, {Da Deppo, V.}, {Debei, S.}, {de Cecco, M.}, {Fulle, M.},
  {Groussin, O.}, {Gutierrez, P. J.}, {Kramm, J.-R.}, {K\"uppers, M.}, {Lara,
  L. M.}, {Lazzarin, M.}, {Lopez Moreno, J. J.}, {Marzari, F.}, {Michalik, H.},
  {Naletto, G.}, {Agarwal, J.}, {G\"uttler, C.}, {Oklay, N.}, \& {Tubiana,
  C.}}]{thomas2015}
{Thomas, N.}, {Davidsson, B.}, {El-Maarry, M. R.}, {et~al.} 2015, A\&A, 583,
  A17

\bibitem[{Tricarico {et~al.}(2021)Tricarico, Scheeres, French, McMahon, Brack,
  Leonard, Antreasian, Chesley, Farnocchia, Takahashi, Mazarico, Rowlands,
  Highsmith, Getzandanner, Moreau, Johnson, Philpott, Bierhaus, Walsh,
  Barnouin, Palmer, Weirich, Gaskell, Daly, Seabrook, Nolan, \&
  Lauretta}]{tricarico2021}
Tricarico, P., Scheeres, D., French, A., {et~al.} 2021, Icarus, 370, 114665

\bibitem[{Tsuchiyama {et~al.}(2011)Tsuchiyama, Uesugi, Matsushima, Michikami,
  Kadono, Nakamura, Uesugi, Nakano, Sandford, Noguchi, Matsumoto, Matsuno,
  Nagano, Imai, Takeuchi, Suzuki, Ogami, Katagiri, Ebihara, Ireland, Kitajima,
  Nagao, Naraoka, Noguchi, Okazaki, Yurimoto, Zolensky, Mukai, Abe, Yada,
  Fujimura, Yoshikawa, \& Kawaguchi}]{tsuchiyama2011}
Tsuchiyama, A., Uesugi, M., Matsushima, T., {et~al.} 2011, Science, 333, 1125

\bibitem[{van~den Wildenberg {et~al.}(2013)van~den Wildenberg, van Loo, \& van
  Hecke}]{wildenberg2013}
van~den Wildenberg, S., van Loo, R., \& van Hecke, M. 2013, Phys. Rev. Lett.,
  111, 218003

\bibitem[{Velick\'y \& Caroli(2002)}]{velicky2002}
Velick\'y, B., \& Caroli, C. 2002, Phys. Rev. E, 65, 021307

\bibitem[{Walsh {et~al.}(2022)Walsh, Ballouz, Jawin, Avdellidou, Barnouin,
  Bennett, Bierhaus, Bos, Cambioni, Connolly, Delbo, DellaGiustina, DeMartini,
  Emery, Golish, Haas, Hergenrother, Ma, Michel, Nolan, Olds, Rozitis,
  Richardson, Rizk, Ryan, Sánchez, Scheeres, Schwartz, Selznick, Zhang, \&
  Lauretta}]{walsh2022-tagsam}
Walsh, K.~J., Ballouz, R.-L., Jawin, E.~R., {et~al.} 2022, Science Advances, 8,
  eabm6229

\bibitem[{Walton(1977)}]{walton1977}
Walton, K. 1977, Geophysical Journal International, 48, 461

\bibitem[{Wang {et~al.}(2020)Wang, Zhang, Feng, Yang, Wu, \& Yue}]{wang2020}
Wang, J., Zhang, M., Feng, L., {et~al.} 2020, Powder Technology, 363, 187

\bibitem[{Watanabe {et~al.}(2017)Watanabe, Tsuda, Yoshikawa, Tanaka, Saiki, \&
  Nakazawa}]{watanabe2017}
Watanabe, S.-i., Tsuda, Y., Yoshikawa, M., {et~al.} 2017, Space Science
  Reviews, 208, 3

\bibitem[{Wright {et~al.}(2020)Wright, Quillen, South, Nelson, S{\'a}nchez,
  Martini, Schwartz, Nakajima, \& Asphaug}]{wright2020}
Wright, E., Quillen, A.~C., South, J., {et~al.} 2020, Icarus, 337, 113424

\bibitem[{Yada {et~al.}(2022)Yada, Abe, Okada, Nakato, Yogata, Miyazaki,
  Hatakeda, Kumagai, Nishimura, Hitomi, Soejima, Yoshitake, Iwamae, Furuya,
  Uesugi, Karouji, Usui, Hayashi, Yamamoto, Fukai, Sugita, Cho, Yumoto, Yabe,
  Bibring, Pilorget, Hamm, Brunetto, Riu, Lourit, Loizeau, Lequertier,
  Moussi-Soffys, Tachibana, Sawada, Okazaki, Takano, Sakamoto, Miura, Yano,
  Ireland, Yamada, Fujimoto, Kitazato, Namiki, Arakawa, Hirata, Yurimoto,
  Nakamura, Noguchi, Yabuta, Naraoka, Ito, Nakamura, Uesugi, Kobayashi,
  Michikami, Kikuchi, Hirata, Ishihara, Matsumoto, Noda, Noguchi, Shimaki,
  Shirai, Ogawa, Wada, Senshu, Yamamoto, Morota, Honda, Honda, Yokota,
  Matsuoka, Sakatani, Tatsumi, Miura, Yamada, Fujii, Hirose, Hosoda, Ikeda,
  Iwata, Kikuchi, Mimasu, Mori, Ogawa, Ono, Shimada, Soldini, Takahashi, Takei,
  Takeuchi, Tsukizaki, Yoshikawa, Terui, Nakazawa, Tanaka, Saiki, Yoshikawa,
  Watanabe, \& Tsuda}]{yada2022}
Yada, T., Abe, M., Okada, T., {et~al.} 2022, Nature Astronomy, 6, 214

\bibitem[{{Yano} {et~al.}(2006){Yano}, {Kubota}, {Miyamoto}, {Okada},
  {Scheeres}, {Takagi}, {Yoshida}, {Abe}, {Abe}, {Barnouin-Jha}, {Fujiwara},
  {Hasegawa}, {Hashimoto}, {Ishiguro}, {Kato}, {Kawaguchi}, {Mukai}, {Saito},
  {Sasaki}, \& {Yoshikawa}}]{science_yano}
{Yano}, H., {Kubota}, T., {Miyamoto}, H., {et~al.} 2006, Science, 312, 1350

\bibitem[{Yu \& Michel(2018)}]{yu2018}
Yu, Y., \& Michel, P. 2018, Icarus, 312, 128

\bibitem[{Zacny {et~al.}(2018)Zacny, Bierhaus, Britt, Clark, Hartzell, Gertsch,
  Kulchitsky, Johnson, Metzger, Reeves, Sanchez, \& Scheeres}]{zacny2018}
Zacny, K., Bierhaus, E.~B., Britt, D.~T., {et~al.} 2018, in Primitive
  Meteorites and Asteroids, ed. N.~Abreu (Elsevier), 439 -- 476

\bibitem[{Zeng {et~al.}(2007)Zeng, Agui, \& Nakagawa}]{zeng2007}
Zeng, X., Agui, J., \& Nakagawa, M. 2007, Journal of Aerospace Engineering, 20,
  116

\bibitem[{Zhai {et~al.}(2020)Zhai, Herbold, \& Hurley}]{zhai2020}
Zhai, C., Herbold, E.~B., \& Hurley, R.~C. 2020, Proceedings of the National
  Academy of Sciences, 117, 16234

\end{thebibliography}
\bibliographystyle{aasjournal}



\end{document}